\documentclass[a4paper,11pt]{article}
\pdfoutput=1 

\usepackage{jheppub} 

\usepackage[T1]{fontenc} 

\title{\boldmath Anisotropic Signatures: Neutrinos - Dark Energy Interaction  and Its Effect on the Transition from Radiation to Matter, and Dark Energy Dominated Phases.}

\author[a,1]{M. Yarahmadi,\note{Also at Some University.}}
\author[a,2]{A. Salehi\note{Corresponding author.}}

\affiliation[a]{Department of Physics, Lorestan University, Khoramabad, Iran}

\emailAdd{yarahmadimohammad10@gmail.com}
\emailAdd{salehi.a@lu.ac.ir}

\abstract{This paper explains the significance of neutrino mass in the cosmic progression from the radiation-dominated phase to matter and subsequently to the dark energy-dominated era. We have put a constraint on the total mass of neutrinos by coupling them with quintessence. For the combination of full data(Pantheon+CMB+BAO+CC), we find $ \sum m_{\nu}<0.101$eV \ \ (95$\% $CL.) and for the relativistic to non-relativistic phase transition redshif  ${z_{\rm nr}} = 180$ which is in the matter-dominated era. Our findings confirm that when neutrinos become non-relativistic, the universe transitions from a radiation-dominated era to a matter-dominated era. Coupled neutrinos with quintessence (CQ) have also a significant impact on transitions from a matter-dominated era to a dark energy era. We have shown this effect by investigating the impact of neutrino mass on the bulk flow direction and amplitude of bulk velocity. Moreover, we have discussed the impact of this coupling on the CMB power spectrum to show the anisotropy in the universe. Finally, we have established a link between the quintessence field coupled with neutrinos and the bulk flow, which allowed us to demonstrate that the mass of neutrinos could be the cause of anisotropy in the universe. 
	}

\begin{document}
	\maketitle
	\flushbottom
	
	\section{Introduction}
Frequent astronomical observations have revealed that the universe is currently undergoing an accelerated expansion phase. (\cite{Ade};\cite{Alam};\cite{Riess1};\cite{Perlmutter}). Today, most cosmologists believe that dark energy is responsible for the accelerated expansion of the universe in the current period.
	The best candidate for dark energy is $\Lambda$CDM model. Despite its considerable success in describing the various available observational data, there are also fundamental problems with the matching model. For example, there is a large discrepancy between observed and predicted values for the vacuum energy density, which is thought to be interpreted as a cosmological constant. Therefore, this problem naturally requires the search for alternatives to the cosmological constant, for where, as in the case of the dark energy equation of state, it differs from the cosmological constant and instead varies with time. Thus, dark energy becomes a dynamic quantity that can in principle help solve the problem of cosmic coincidence or fine-tune the initial conditions. The goal of particle physics is to provide a phenomenological description of dark energy first and then, if possible, relate this effective solution to some aspect of high-energy physics. If dark energy is considered a dynamically evolving entity, the simplest theoretical explanation for it is obtained by introducing a conventional scalar field, i.e. a scalar particle. Dark energy models built from the conventional scalar field are collectively known as quintessence. 
	In the framework of the standard model of particle physics, neutrinos are massless particles belonging to a family of fundamental particles called leptons. In beta decay, the neutron turns into a proton, an electron, and an antineutrino, which was unknown at that time. About 25 years later, the first experimental detection of neutrinos occurred, which is now known as the Kuan-Rhines neutrino experiment \cite{Cowan}. In this experiment, neutrinos created in beta decay interact with protons. In 1962, Leon Lederman, Melvin Schwartz, and Jack Steinberger identified the muon neutrino and showed that there is more than one type of neutrino \cite{Danby}. In the year 2000, the tau neutrino was discovered for the first time by the donut experiment at the Fermi laboratory \cite{Kodama}. In the mid-1960s, in the famous Homestake experiment conducted by the American scientist Raymond Davis Jr., it was observed that the number of electron neutrinos that reach the earth from the sun is much less than predicted. This problem is called the solar neutrino problem. It can remain unsolved for 30 years. However, this issue can be explained by the mechanism that Bruno Ponte Coro proposed for the first time in 1957 \cite{Pontecorvo}. Ponte Coro proposed that if neutrinos had mass, the neutrino flavor would change through a mechanism called neutrino flavor oscillations. Therefore, if neutrinos have mass, the cosmic neutrino problem is solved. In 1988, the first experimental discovery of neutrino oscillations was made by the Super Kamiokande experiment \cite{Fukuda}. This important experiment and several other experiments that were performed later (for example, refer to sources \cite{Ahma} and \cite{Eguchi}) show that neutrinos are particles with mass. While in the framework of the standard model of particle physics, neutrinos are massless. Most of the neutrinos that reach the earth are from the sun. Neutrinos produced in solar processes are electron neutrinos. Using measurements of oscillation experiments, it has been determined that at least two of the three special types of neutrino mass states have a non-zero mass \cite{Ahmad2},\cite{Davis1},\cite{Fukuda},\cite{Kajita}. However, many questions about neutrinos
	remain open until today. For example, the exact neutrino mass ordering (neutrino hierarchy) remains
	unknown, as well as the absolute neutrino masses. It is also unclear, there are only three neutrino species or additional “sterile” neutrinos. Besides, the Standard Model of Particle Physics does not explain yet how the neutrino masses are generated. Cosmology has helped to come
	closer to answering some of these questions. In particular, cosmological observations are sensitive
	to the imprint of neutrinos on structure formation, because neutrinos suppress structure formation
	on small scales and slow down the growth of structure at all scales. There are more
	than 300 neutrinos for every $cm^{3}$ in the universe! 
	
	In cosmological studies, the "bulk flow" refers to the large-scale coherent motion of galaxies or galaxy clusters in a particular direction, suggesting the presence of unaccounted-for dynamics on cosmological scales. According to the standard cosmological model, the movement of galaxy clusters should be randomly distributed in all directions due to cosmic background radiation. During the 1970s, a fascinating phenomenon known as bulk flow was discovered. It was noticed that galaxies in specific regions of the sky were moving in a particular direction, indicating an organized motion pattern \cite{Conklin}\cite{Henry}.
	It was suspected that the gravitational attraction towards a nearby
	over density might be responsible for the bulk motion. At first, it seemed that the cause of the bulk flow was the Virgo supercluster, which is located at a distance of 17 megaparsecs from the center of the Milky Way.  However, by analyzing three years of WMAP data using kinematics by Sunaev-Zeldevich, astronomers: Alexander Kashlinsky, Arturo Barandele-David Kochesky, and Abel Ebling,\cite{Kashlinsky1};\cite{Kashlinsky3};\cite{Kashlinsky4} evidence of a surprising homogeneity of the flow of galaxy clusters between the constellations of Centauri and Vela. They found that they are moving at a velocity of 600 to 1000 kilometers per second \cite{Kashlinsky11}. The researchers suggested that this movement may be a remnant of the influence of invisible regions from the pre-inflationary universe. Telescopes cannot see events before about 380,000 years after the Big Bang when the universe becomes clear (cosmic background). This corresponds to the particle horizon at a distance of about 46 billion light-years (\cite{Barandela};\cite{Kashlinsky22};\cite{Bonvin};\cite{Colin}). Research conducted in 2013 by the Planck Space Telescope revealed no evidence of bulk flow on a large scale. This effectively contradicted claims of gravitational effects beyond the observable universe or the existence of multiverses. However, in 2015, Kashlinsky and colleagues \cite{Atrio-Barandela}  claimed to have observed evidence of dark flow, utilizing data from Planck and WMAP. Specifically, the dark flow was noted to be in the direction of constellations Centaurus and Hydra, which aligns with the direction of the Great Attractors. Nonetheless, the source of the large absorption appeared to be from a massive cluster of galaxies referred to as the Norma cluster, which is located 250 million light-years away from the Milky Way. Further research conducted by Kashlinsky and colleagues in March 2010 extended their earlier findings, using 5 years of WMAP data and examining 700 galaxy clusters. The team also presented a list of galaxy clusters based on different distances, with four parts. They then investigated the preferred flow direction for the clusters in each section. The group has cataloged this dark halo effect as far as 2.5 billion light-years and expects to expand the catalog to twice the current distance.

	In this paper, we first put constraints on the total mass of neutrinos by the interaction between neutrinos with quintessence scalar field; then calculate the redshift at which  $\sum m_{\nu}$ will become non-relativistic (${z_{\rm nr}}$) to investigate the effect of neutrinos on evolution from radiation dominated to matter dominated. Moreover, we investigate the effect of neutrinos on evolution from radiation-dominated to dark energy-dominated. Furthermore, we show that when considering the interaction between neutrinos and the quintessence field, the direction of bulk flow is very close to the direction of dark energy dipole beyond the local universe.

	\section{Quintessence scalar field coupled with neutrino}
	Quintessence models often feature tracker solutions, meaning that their energy density naturally converges to a specific value over time, irrespective of the initial conditions. This tracker behavior helps to avoid fine-tuning issues and ensures that quintessence becomes dynamically significant during the relevant cosmic epochs. Dark energy, in the form of quintessence, undergoes a dynamic evolution over time. When coupled with neutrinos, it creates a scenario where both the dark energy and neutrinos adapt to cosmic changes in a dynamic manner. This dynamic evolution may offer a more natural explanation for the observed cosmic acceleration and the properties of neutrinos.
	The coupling of neutrinos to a dynamic quintessence field introduces an additional layer of complexity to the cosmic fluid. This complexity can lead to directional variations in the expansion rate and large-scale structure formation. Studying this model can help cosmologists understand and potentially explain observed anisotropies in the universe.
	
	The simplest scalar field that can explain dark energy is Quintessence which was ﬁrst proposed as an alternative to the cosmological constant in 1998 (\cite{Caldwell}). Although applications of the scalar field in late cosmology were widely studied even before 1998 (\cite{Ratra} \cite{Wetterich}). The first example of this scenario was proposed by \cite{Ratra} \cite{Wetterich}. It has been proposed by some physicists as the fifth fundamental force. \cite{Carroll},\cite{Wetterich},\cite{Cicoli}. The energy density of quintessence does not need to be very small with respect to radiation or matter in the early Universe, unlike the cosmological constant scenario. 
	The action that represents our physical system is:
	\begin{equation}
		\begin{split}
			S = \int {{d^4}x} \sqrt { - g} (\frac{R}{{2{\kappa ^2}}} + {L_m} + {L_\sigma }) , \\
			{L_\sigma } = \frac{{ - 1}}{2}{g^{\mu \nu }}{\partial _\mu }\sigma {\partial _\nu }\sigma  - V(\sigma ),
		\end{split}
	\end{equation}
	where $L_{\sigma}$ is the Lagrangian and  V($\sigma$) is a general self-coupling potential $\sigma$ for which must be positive for physically acceptable fields.
	
	We work in flat universe and using the FLRW metric we have:
	
	\begin{equation}
		d{s^2} = - d{t^2}+a^{2}(t)(dx^{2}+dy^{2}+dz^{2}).
	\end{equation}
	
	 Einstein's field equations  are reduced to the following Friedman equations and acceleration :
	\begin{equation}
		3{H^2} = {\kappa ^2}(\rho_{m}   + \frac{1}{2}{{\dot \sigma }^2} + V),
	\end{equation}

	\begin{equation}
		2\dot H + 3{H^2} =  - {\kappa ^2}(\omega \rho_{m}  + \frac{1}{2}{{\dot \sigma }^2} + V).
	\end{equation}
	While the Klein-Gordon equation  is simplified as follows:

	\begin{equation}
		(\ddot \sigma  + 3H{{\dot \sigma }^2} + {V_{,\sigma }}) = 0.
	\end{equation}
	The energy density and pressure of the scalar field are defined as follows:
	
	\begin{equation}
		{\rho _\sigma } = \frac{1}{2}{{\dot \sigma }^2} + V,
	\end{equation}
	\begin{equation}
		{P_\sigma } = \frac{1}{2}{{\dot \sigma }^2} - V.
	\end{equation}
	
	Therefore, the equation of state is
	
	\begin{equation}
		{\omega _\sigma } = \frac{{{P_\sigma }}}{{{\rho _\sigma }}} = \frac{{\frac{1}{2}{{\dot \sigma }^2} - V}}{{\frac{1}{2}{{\dot \sigma }^2} + V}}.
	\end{equation}
	Note that $\omega_{\sigma}$ is a dynamically evolving parameter that can take values in the range [-1, 1].
	
	Cosmological models usually have non-linear differential equations and it is not always possible to find exact solutions for a non-linear system. The method of dynamical systems that studies nonlinear systems can help to investigate the qualitative behavior of these systems. Usually, the new dimensionless variables are defined as normalized ones so that an independent system is obtained. These variables are directly related to visible physical parameters. By finding the stationary points and their stability conditions, it is possible to qualitatively study the beginning and the end of the probability of the Universe. This type of analysis is not a new topic in general relativity and cosmology. Almost all the important models of general relativity and cosmology have been investigated by the method of dynamical systems.

	In the following, we discrete the $ \rho_{m} $ and $ p_{m} $ to ${\rho _m} = {\rho _b} + {\rho _c} + {\rho _\nu } $ and $ {p_m} = {p_b} + {p_c} + {p_\nu } $ in field equations (2.3) and (2.4) and solving the equations with dynamical systems, we will investigate the evolution of deceleration parameter. Einstein's field equations  are reduced to the following Friedman equations and acceleration :
	\begin{equation}\label{fried}
		\begin{split}
			3H^{2}=\kappa^{2}\left(\rho_{\rm b}+\rho_{\rm c}+\rho_{\rm r}+\rho_{\nu}-\frac{1}{2}\dot{\sigma}^{2}+V(\sigma)\right),
		\end{split}
	\end{equation}
	where $\rho_{\rm b}$ is baryon density and $\rho_{\nu}$ is density of neutrino, $\rho_{\rm c}$ is cold dark matter density.
	\begin{equation}\label{fried}
		\begin{split}
			2\dot{H}+3H^{2}=\kappa^{2}(-\omega_{\rm b}\rho_{\rm b}+\omega_{\rm c}\rho_{\rm c}+\omega_{\rm r}\rho_{\rm r}+\omega_{\nu}\rho_{\nu}+\frac{1}{2}\dot{\sigma}^{2}+V(\sigma)).
		\end{split}
	\end{equation}
	The energy density conservation equations are:
	\begin{equation}\label{fried}
		\begin{split}
			\dot{\rho}_{\sigma}+3H\rho_{\sigma}(1+\omega_{\sigma})=-\alpha\rho_{\nu}(1-3\omega_{\nu})\dot{\sigma}.
		\end{split}
	\end{equation}
	In this paper we follow the idea that proposed by  by\cite{Brookfield}. In the cosmological context, neutrinos cannot be described
	as a fluid. Instead, one must solve the distribution
	function $f(x^{i}, p^{i}, \tau)$ in phase space (where $\tau$ is the
	conformal time). Considering the case that neutrinos
	are cohesionless, the distribution function
	$f$ does not depend explicitly on time. Solving the Boltzmann
	equation, one can then calculate the energy density
	stored in neutrinos ($f_{0}$ is the background neutrino distribution
	function):
	\begin{equation}
		\begin{split}
			\rho_{\nu}=a^{-4}\int q^{2}\sqrt{q^{2}+m_{\nu}(\phi)^{2}a^{2}}f_{0}(q)dqd\Omega
		\end{split}
	\end{equation}
	\begin{equation}
		\begin{split}
			p_{\nu}=\frac{1}{3}a^{-4}\int \frac{q^{2}}{\sqrt{q^{2}+m_{\nu}(\phi)^{2}a^{2}}}f_{0}(q)dqd\Omega
		\end{split}
	\end{equation}
	The evolution equation for its energy density according to \cite{Brookfield}
	\begin{equation}\label{fried}
		\begin{split}
			\dot{\rho}_{\nu}+3H\rho_{\nu}(1+\omega_{\nu})=\alpha\rho_{\nu}(1-3\omega_{\nu})\dot{\sigma},
		\end{split}
	\end{equation}
	
	where $\alpha$ denotes coupling constant which can be related neutrino mass $m_{\nu}$ with relation $\alpha=\frac{d\ln m_{\nu}}{d\sigma}$ and $m_{\nu}=m_{0}exp(\frac{\alpha \sigma }{m_{pl}})$.
	
	Note that if $\omega_{\nu} = \frac{1}{3}$, the coupling
	vanishes making neutrinos non-interacting particles.

	\begin{equation}\label{fried}
		\begin{split}
			\dot{\rho}_{\rm c}+3H\rho_{\rm c}=-\beta \rho_{\rm c}\dot{\sigma},
		\end{split}
	\end{equation}
	also
	\begin{equation}\label{fried}
		\begin{split}
			\dot{\rho}_{\rm b}+3H\rho_{\rm b}=0,
		\end{split}
	\end{equation}
	and
	\begin{equation}\label{fried}
		\begin{split}
			\dot{\rho}_{\rm r}+4H\rho_{\rm r}=0,
		\end{split}
	\end{equation}
	respectively.

	The coupling effect becomes significant only in the case of non-relativistic neutrinos. This is because relativistic neutrinos exhibit an approximate relationship, $P_{\upsilon} \approx \frac{\rho_{\upsilon}}{3}$, which renders both sides of Eqs. (2.11) and (2.14) negligible. Consequently, the standard, uncoupled conservation equations are reinstated.

	In addition, from the resulting of
	equations of the scalar field :
	\begin{equation}\label{fried}
		\begin{split}
			\ddot \sigma  + 3H\dot \sigma  + \frac{{dV}}{{d\phi }} + \alpha \dot \sigma (1 - {\omega _\nu }){\rho _\nu } - \beta {\rho _c}\dot \sigma  = 0
		\end{split}
	\end{equation}
	where as before $\dot{a}$  means differentiation with respect to the coordinate time t, $ \alpha $ is coupling constant between dark energy and neutrinos  and $ \beta $ is coupling constant between dark energy and cold dark matter. 
	We have derived the coupling constant, denoted as $\alpha$, characterizing the interaction between neutrinos and the quintessence scalar field. Our findings indicate that higher values of $\alpha$ generally correspond to increased neutrino masses within the cosmic framework.
	If we considered the standard model, the right side of the equation would be zero, and the neutrinos are effectively uncoupled.
For most of the Universe's history, the neutrinos are
highly relativistic and $(\rho_{\nu}-3p_{\nu})\approx0$ such that the scalar
field and the neutrinos are effectively uncoupled, here only the coupling parameter $\beta$ is important. After the neutrinos become non-relativistic hence the coupling parameter $\alpha$ also becomes important. We consider an exponential potential $V=M_{pl}^{4}\exp(-\lambda\frac{\sigma}{M_{pl}})$, where $\lambda$ is a dimensionless parameter
that determines the slope of the potential. The motivation for choosing these functions have been investigated in \cite{Wetterich}.

In order to simplify the field equations, we introduce the following new variables,
	\begin{equation}\label{fried}
		\begin{split}
			\Omega_{\rm b}=\frac{\kappa^{2}\rho_{\rm b}}{3H^{2}}, \ \ \ \ \Omega_{\nu}=\frac{\kappa^{2}\rho_{\nu}}{3H^{2}}, \ \ \ \ \Omega_{\rm r}=\frac{\kappa^{2}\rho_{\rm r}}{3H^{2}},\\ \ \ \ \Omega_{\rm c}=\frac{\kappa^{2}\rho_{c}}{3H^{2}}, \ \ \ \ 
			\Omega_{\sigma}=-\frac{\kappa\dot{\sigma}}{\sqrt{6}H}, \ \ \
			\Omega_{\rm V}=\frac{\kappa^{2}V(\sigma)}{3H^{2}},
		\end{split}
	\end{equation}
	were $ \Omega_{\rm b} $ is baryon density, $ \Omega_{\rm r} $ is radiation density, $ \Omega_{\rm c} $ is cold dark matter density, $ \Omega_{\nu} $ is neutrino density and $ \Omega_{\sigma} $ and $ \Omega_{\rm V} $ denotes as dark energy density.
	In term of new variable the Friedmann equation  puts a constraint on new variables as
	\begin{equation}\label{cons}
		\begin{split}
			\Omega_{\rm V}=1-\Omega_{\rm b}-\Omega_{\nu}-\Omega_{\rm r}-\Omega_{\rm c}-\Omega_{\sigma}^{2}.
		\end{split}
	\end{equation}
	Therefore, the equations are simplified as follows:
	\begin{equation}\label{fried}
		\frac{d\Omega_{\rm b}}{dN}=-3\Omega_{\rm b}-2\Omega_{\rm b}\frac{\dot{H}}{H^{2}}, \ \ \ \ \ \
	\end{equation}
	\begin{equation}\label{fried}
		\frac{d\Omega_{\nu}}{dN}=-\Omega_{\nu}\left(3(1+\omega_\nu)+\sqrt{6}\Omega_{\sigma}\beta(1-3\omega_\nu)\right)-2\Omega_{\nu}\frac{\dot{H}}{H^{2}}, \ \ \
	\end{equation}
	\begin{equation}\label{fried}
		\frac{d\Omega_{\rm r}}{dN}=-4\Omega_{\rm r}-2\Omega_{\rm r}\frac{\dot{H}}{H^{2}},\ \ \
	\end{equation}
	\begin{equation}\label{fried}
		\frac{d\Omega_{\rm c}}{dN}=-\Omega_{\rm c}\left(3-\sqrt{6}\Omega_{\sigma}\alpha+2\frac{\dot{H}}{H^{2}}\right),
	\end{equation}
	\begin{equation}\label{fried}
		\begin{split}	
			\frac{{d\Omega_{\sigma}}}{{dN}} =  \frac{{3\lambda {\Omega_{\rm V}}}}{{\sqrt 6 }} - \frac{{3\beta {\Omega_{\nu}}(1 - 3{\omega _\nu })}}{{\sqrt 6 }} + \frac{{9\alpha }}{{\sqrt 6 }}{\Omega_{\rm c}} + 3{\Omega_{\sigma}} - {\Omega_{\sigma}}\frac{{\dot H}}{{{H^2}}}
		\end{split}
	\end{equation}
	
	Following the reference \cite{Wali}, we shall
	use the following ansatz for $\omega_{\nu}(z)$
	\begin{equation}
		\omega_{\nu}(z)=\frac{p_{\nu}}{\rho_{\nu}}=\left(1+tanh(\frac{ln(1+z)-z_{\rm eq}}{z_{\rm dur}})\right)
	\end{equation}
	we can drive
	
	\begin{equation}\label{fried}
		\frac{d\omega_{\nu}}{dN}=\frac{2\omega_{\nu}}{z_{\rm dur}}(3\omega_{\nu}-1),
	\end{equation}
	where $z_{eq}$ determines the transition redshift where matter and radiation energy densities become equal and $z_{\rm dur}$ determines how fast this transition is realized and
	$N=\ln a$. In term of the new dynamical variable, we also have,
	\begin{equation}\label{fried}
		\begin{split} \frac{\dot{H}}{H^{2}}=\frac{1}{2}\left(-3-\Omega_{\rm r}-3\omega_{\nu}\Omega_{\nu}+3\Omega_{\sigma}^{2}+3\Omega_{\rm V}\right).
		\end{split}
	\end{equation}
	
	The parameter mentioned above holds significant importance in cosmology, as it directly influences essential cosmological parameters. For instance, it plays a crucial role in determining the deceleration parameter ($q$) and the effective equation of state ($w_{\rm eff}$), both of which are fundamental in understanding the dynamics of the universe.
	
	The deceleration parameter $q$ is defined as $q = -1 - \frac{\dot{H}}{H^2}$, where $\dot{H}$ represents the time derivative of the Hubble parameter $H$. This parameter quantifies whether the universe is currently experiencing cosmic acceleration ($q < 0$) or cosmic deceleration ($q > 0$).
	
	Similarly, the effective equation of state ($w_{\rm eff}$) is expressed as $w_{\rm eff} = -1 - \frac{2}{3}\frac{\dot{H}}{H^2}$. It characterizes the nature of cosmic expansion, providing insights into whether the universe's expansion is dominated by conventional matter ($w_{\rm eff} \approx 0$), dark energy ($w_{\rm eff} < -1/3$), or a cosmological constant ($w_{\rm eff} = -1$).
	
	Moreover, this parameter also plays a pivotal role in calculations related to the luminosity distance, which is crucial in cosmological observations, particularly in the context of measuring the distance to celestial objects and understanding the universe's expansion history.

	\section{Numerical analysis}
	In this paper we use the Markov Chain Monte Carlo (MCMC) method to best fit parameters.
	In cosmology, the (MCMC) method plays a crucial role in addressing complex statistical and computational challenges inherent in understanding the universe's large-scale structure, dynamics, and the properties of cosmic components. Cosmological models often involve numerous parameters that describe the properties of the universe, such as the density of dark matter, dark energy, and the Hubble constant.  MCMC is employed for fitting cosmological models to observational data. Also, dynamical systems serve as the mathematical backbone for understanding how components like dark matter and dark energy evolve. MCMC steps in to tackle the challenge of exploring these intricate parameter spaces and pinpointing the best-fit models based on observational data. This collaboration is pivotal in deciphering cosmic complexities, addressing nonlinearities, and dealing with uncertainties. In our research, we employ the CAMB code to thoroughly study complex cosmological phenomena. We pay particular attention to the impact of coupled neutrinos on the evolution of the universe. CAMB's flexibility enables us to investigate the intricate relationship between cosmological parameters and gain a comprehensive understanding of the fundamental physics that shapes the universe's evolution. In this paper, we use the Pantheon + Analysis (\cite{Scolnic}) of 1071 supernovae covering the redshift $0.001 < z < 2.3$. Pantheon+ is an updated version of the original Pantheon analysis (Scolnic et al. 2018b). It utilizes the analysis framework of the original Pantheon and incorporates a larger number of SN Ia samples. Additionally, it includes samples that are in galaxies with measured Cepheid distances. This allows for the simultaneous constraint of parameters that describe the complete expansion history and the local expansion rate ($ H_0 $). Also, we used CMB data:
	We used the latest large-scale cosmic microwave background (CMB) temperature and
	polarization angular power spectra from the final release of Planck 2018 plikTTTEEE+lowl+lowE
	\cite{43}. 
	BAO data:
	We also used the various measurements of the Baryon Acoustic Oscillations (BAO) from:\cite{Ratra1}
\begin{table}
	\centering
		\caption{12 BAO data.}\label{tab:bao}
		\setlength{\tabcolsep}{2.3mm}{
			\begin{tabular}{lccc}
				\hline
				$z$ & Measurement{a} & Value \\
				\hline
				\hline
				$0.122$ & $D_V\left(r_{s,{\rm fid}}/r_s\right)$ & $539\pm17$ \\
				$0.38$ & $D_M/r_s$ & 10.23406 \\
				$0.38$ & $D_H/r_s$ & 24.98058 \\
				$0.51$ & $D_M/r_s$ & 13.36595 \\
				$0.51$ & $D_H/r_s$ & 22.31656 \\
				$0.698$ & $D_M/r_s$ & 17.85823691865007 \\
				$0.698$ & $D_H/r_s$ & 19.32575373059217 \\
				$0.835$ & $D_M/r_s$ & $18.92\pm0.51$ \\
				$1.48$ & $D_M/r_s$ & 30.6876 \\
				$1.48$ & $D_H/r_s$ & 13.2609 \\
				$2.334$ & $D_M/r_s$ & 37.5 \\
				$2.334$ & $D_H/r_s$ & 8.99 \\
				\hline
		\end{tabular}}
\end{table}
were $D_V$, $r_s$, $r_{s, {\rm fid}}$, $D_M$, $D_H$, and $D_A$ have units of Mpc.

 	$\bullet$$\bullet$The 12 BAO measurements listed in Table \ref{tab:bao} cover the redshift range $0.122 \leq z \leq 2.334$. The quantities listed in Table 1 are described as follows:

$\bullet$ $D_V(z)$: Spherically averaged BAO distance, $D_V(z)=[czH(z)^{-1}D^2_M(z)]^{1/3}$, where $c$ is the speed of light and the angular diameter distance $D_A(z) = D_M(z)/(1+z)$ with $D_M(z)$ defined in the following\\
$\bullet$ $D_H(z)$: Hubble distance, $D_H(z)=c/H(z)$\\
$\bullet$ $r_s$: Sound horizon at the drag epoch, $r_{s, {\rm fid}}=147.5$ Mpc.\\
$\bullet$ $D_M(z)$: Transverse comoving distance,
		\begin{equation}
			\label{eq:DM}
			D_M(z) = 	D_C(z) \ \ \ \  \text{if}\ \Omega_{k0} = 0,\\
		\end{equation}
		where the comoving distance
		\begin{equation}
			\label{eq:gz}
			D_C(z) = c\int^z_0 \frac{dz'}{H(z')}.
		\end{equation}

	CC data: The cosmic chronometer (CC) data covering the redshift $0.07 < z < 2.5$.(Table 1)\\
	\begin{table*}
		\centering
		\scriptsize
		\caption{32 $H(z)$ (CC) data.}\label{tab:hz}
		\setlength{\tabcolsep}{7.5mm}{
			\begin{tabular}{lcc}
				\hline
				$z$ & $H(z)$ & $\sigma$\\
				\hline
				0.07 & $69.0$ & 19.6\\
				0.09 & $69.0$ & 12.0\\
				0.12 & $68.6$ & 26.2\\
				0.17 & $83.0$ & 8.0\\
				0.2 & $72.9$ & 29.6\\
				0.27 & $77.0$ & 14.0\\
				0.28 & $88.8$ & 36.6\\
				0.4 & $95.0$ & 17.0\\
				0.47 & $89.0$ & 50.0\\
				0.48 & $97.0$ & 62.0\\
				0.75 & $98.8$ & 33.6\\
				0.88 & $90.0$ & 40.0\\
				0.9 & $117.0$ & 23.0\\
				1.3 & $168.0$ & 17.0\\
				1.43 & $177.0$ & 18.0\\
				1.53 & $140.0$ & 14.0\\
				1.75 & $202.0$ & 40.0\\
				0.1791 & 74.91 & 4.00\\
				0.1993 & 74.96 & 5.00\\
				0.3519 & 82.78 & 14\\
				0.3802 & 83.0 &  13.5\\
				0.4004 & 76.97 &  10.2\\
				0.4247 & 87.08 &  11.2\\
				0.4497 & 92.78 &  12.9\\
				0.4783 & 80.91 &  9\\
				0.5929 & 103.8 & 13\\
				0.6797 & 91.6 & 8\\
				0.7812 & 104.5 & 12\\
				0.8754 & 125.1 & 17\\
				1.037 & 153.7 & 20\\
				1.363 & 160.0 & 33.6\\
				1.965 & 186.5 & 50.4\\
				\hline
		\end{tabular}}
	\end{table*}

	\section{From radiation dominated to dark energy dominated}
	The combined effects of coupled quintessence with neutrinos can impact the evolution of the universe on large scales, affecting the growth of structures,\cite{Yarahmadi} the expansion rate, and the distribution of matter. These effects can potentially be probed through observational data, such as measurements of large-scale structures, the cosmic microwave background, and the clustering of neutrinos\cite{Yarahmadi}. When we talk about coupling neutrinos to quintessence, we are essentially talking about the interactions between neutrinos and the evolving scalar field associated with quintessence. This coupling mechanism can impact the behavior of both neutrinos and the scalar field, which in turn can affect the overall dynamics of the universe\cite{Yarahmadi1}. As neutrinos gradually decelerate and become non-relativistic, their role in the evolution of the Universe becomes increasingly significant. The total mass of neutrinos refers to the amount of energy that is associated with the background of ancient neutrinos that were left over from the Big Bang. These neutrinos initially existed in equilibrium with the early cosmic plasma but eventually separated, approximately one second after the Big Bang, due to their weak interactions. Despite later interactions involving electrons and positrons influencing the distribution of photons, we can still connect the temperatures of these two groups of particles by examining their entropy. From this, we can relate the cosmic neutrino density to the sum of the individual mass eigenstates, which is given by $\sum m_{\nu}$.
	\begin{equation}\label{fried}
		{\Omega _\nu } = \frac{{\sum {{m_\nu }} }}{{93.14{h^2}}}
	\end{equation}
	This relation helps us to obtain the sum of the neutrino masses and the absolute scale.
	The above relation immediately enables an upper bound of $\sum m_{\nu}$.
	
	In Eq. (2.19), if we best fit  value of $ \Omega_{\nu}$ and the value of h, then automatically we can obtain the total mass of neutrinos.  The more direct effects of the neutrino depend on whether they are relativistic, non-relativistic
	and also over what scale one is considering. In the early Universe these particles will naturally
	behave like radiation and at some point, depending on their mass, will make a transition to become
	matter-like. Specifically, the massive neutrino species start to become non-relativistic at a redshift
	given by,
	\begin{equation}\label{fried}
		1+z_{nr}\approx\frac{2}{3}\times10^{3}(\frac{\sum m_{\nu}}{ev})
	\end{equation}
	
For a significant duration of the Universe’s history, neutrinos exhibit high relativistic behavior, leading to the effective uncoupling of the scalar field and neutrinos, with the coupling parameter $\beta$ assuming sole significance. Subsequent to the transition of neutrinos into a non-relativistic state, the coupling parameter $\alpha$ becomes a crucial factor. Confining our focus to parameter $\alpha$, the following constraints have been derived: $\alpha = 5.987^{+1.68}_{-1.27}, \alpha = 5.64^{+1.6}_{-1.3}, \alpha = 5.27^{+1.36}_{-1.23}, \alpha = 5.44^{+1.2}_{-1.1}$, at $68\%$ CL for Pantheon+, CC, CMB + BAO, and Pantheon + CC + CMB + BAO, respectively.

To evoke growing neutrino quintessence, certain conditions must be satisfied:
\begin{itemize}
	\item $V(\sigma)$ must exhibit a negative gradient, inducing an increase in the scalar field value over time. This gradient must be steep enough for $\phi$ to attain sufficiently large values in the late Universe, functioning as dark energy.
	\item $|\alpha|$ must be adequately large when neutrinos transition to a non-relativistic state, enabling $\beta(\rho_\nu - 3p_\nu)$ to act as a robust restoring force, halting the evolution of $\sigma$ as per Eq. (2.18).
\end{itemize}

The best-fitted values of $(\alpha, \lambda)$ adhere to the aforementioned conditions. Despite the seemingly modest value of $\nu$, the role of neutrinos remains crucial in shaping the cosmos' evolution, attributed to their substantial coupling represented by $\alpha$. We listed another bestfitted cosmological parameter in table 3. 

	In the following, we put constraint on total mass of neutrino by coupling the neutrino with scalar field and we obtained the neutrino mass for each data as follows:
	
	From the analysis of the CMB + BAO data and, we find that
	$\sum m_{\nu}<0.115$eV \ \ (95$\% $CL.)  and using Pantheon+ we find $ \sum m_{\nu}<0.201$eV \ \ (95$\% $CL.),and using CC data we find $ \sum m_{\nu}<0.16$eV \ \ (95$\% $CL.)  and for combination of full data(Pantheon+CMB+BAO+CC) we find $ \sum m_{\nu}<0.101$eV \ \ (95$\% $CL.).\\ Observational constraints at (95$\% $CL.) on cosmological parameters are in Table 3.

	Cosmology is sensitive the total mass of neutrino. According to the above equation and the values obtained for the neutrino mass in quintessence model, we will calculate the  redshift at which  $\sum m_{\nu}$ will become non-relativistic ${z_{\rm nr}}$. We found:\\
	
	$\bullet$  For the (combination data)  we obtained  ${z_{\rm nr}} = 180$
	
	$\bullet$  For the Pantheon+ data   we obtained  ${z_{\rm nr}} =370 $
	
	$\bullet$  For the CMB + BAO   we obtained  ${z_{\rm nr}} =210 $
	
	$\bullet$  For the CC data  we obtained  ${z_{\rm nr}} = 302$\\

which is in the region that enters the matter-dominated era(According Fig. 6). This results are in abroad with \cite{Yarahmadi}.

\begin{table*}	
	\caption{: Comparison between   $H_{0}$,   $\alpha$,   $\beta$,  $\lambda$ for different combination of data set.
	} 
	\centering 
	\begin{tabular}{c@{\hspace{2mm}} c@{\hspace{2mm}} c@{\hspace{2mm}} c@{\hspace{2mm}} c@{\hspace{2mm}}
			c@{\hspace{2mm}} c@{\hspace{2mm}} c@{\hspace{2mm}} c@{\hspace{2mm}}  c@{\hspace{2mm}}
			c@{\hspace{2mm}} c@{\hspace{2mm}}c@{\hspace{2mm}}c@{\hspace{2mm}}c@{\hspace{2mm}}} 
		\hline\hline 
		Dataset   & $H_{0}$ &  $\alpha$ &  $\beta$ & $\lambda$&\\
		\hline 
		CMB+BAO  & $69.18\pm 1.8 $ &  $5.27^{+1.36}_{-1.23}$ &  $-4.09^{+1.6}_{-1.54}$ & $8.22\pm0.7$ & \\
		\hline 
		Pantheon+& $71.74\pm1.4$ &  $5.987^{+1.68}_{-1.27}$ &  $-2.97^{+1.8}_{-1.7}$ & $8.014^{+0.8}_{-0.75}$ & \\
			\hline 
		CC& $70.61\pm1.4$ &  $5.64^{+1.6}_{-1.3}$ &  $-3.24^{+1.73}_{-1.89}$ & $6.714^{+0.9}_{-0.9}$ & \\
		\hline 
		CMB+BAO+Pantheon+CC & $68.83\pm1.43$ &  $5.44^{+1.2}_{-1.1}$ &  $-4.17^{+1.3}_{-1.6}$ & $8.1^{+0.74}_{-0.63}$ &  \\
		\hline 
	\end{tabular}
\end{table*}

The mass of neutrinos is a critical factor in the history of the Universe's expansion and the growth of cosmic ray disturbances. It also affects the anisotropy spectrum of cosmic radiation and the observations of large-scale structures. The inclusion of coupled quintessence in cosmological models can lead to several changes in the Cosmic Microwave Background (CMB) power spectrum compared to the standard $\Lambda $CDM  model. 
	
	\begin{figure}
		\centering
		\includegraphics[width=15 cm]{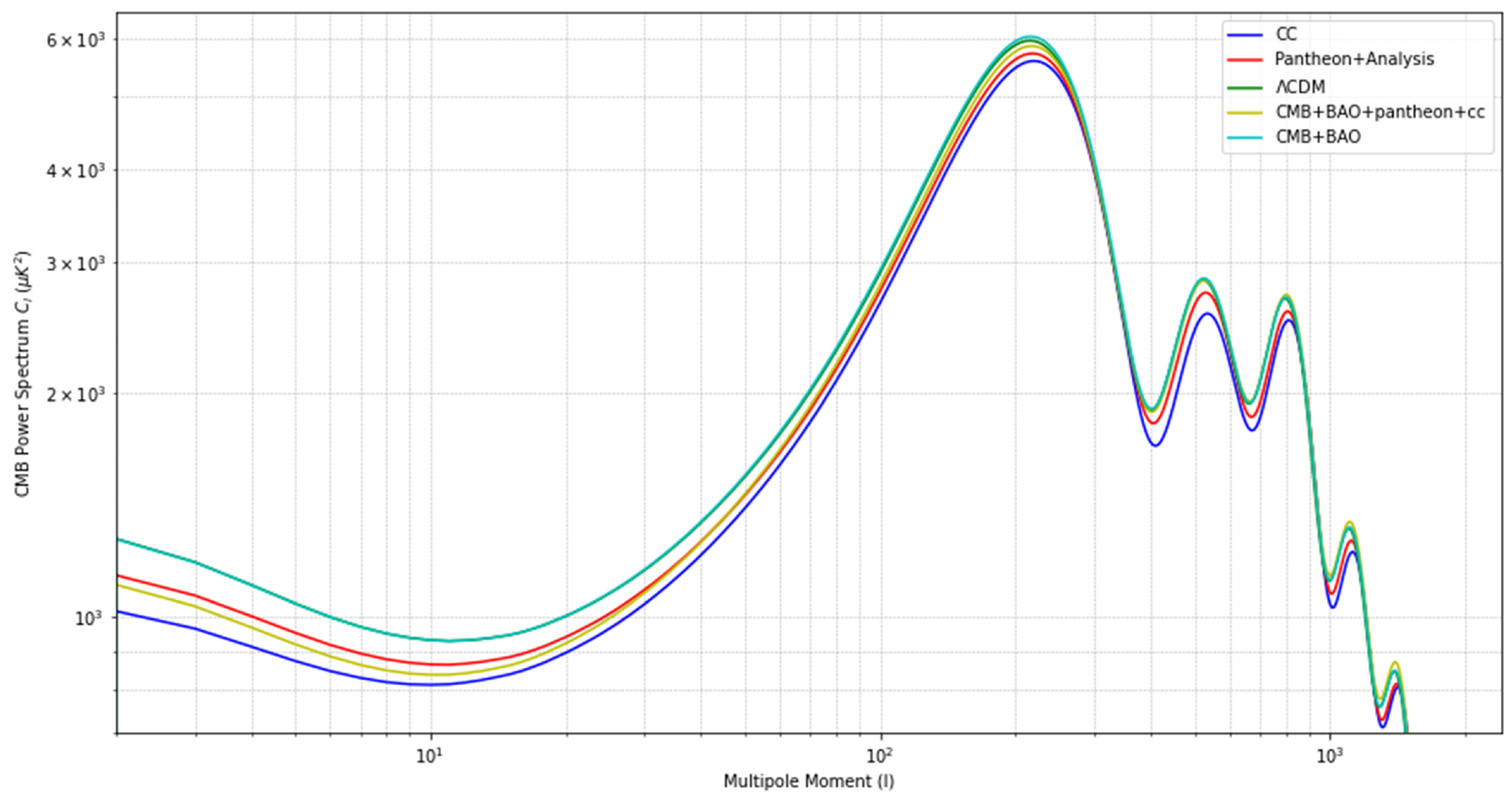}
		\vspace{-0.12cm}
		\caption{\small{Comparison the CMB power spectrum  between coupled quintessence in different combination of data sets  and $\Lambda $CDM model.}}\label{fig:omegam2}
	\end{figure}
The Sachs-Wolfe effect originates from the interaction between gravitational potentials and photons as they traverse through evolving gravitational fields in an expanding universe.The study of coupled neutrinos with quintessence reveals that gravitational fields undergo changes that leave an imprint on the CMB power spectrum. Figure 1 illustrates the Sachs-Wolfe effect for $l=10$ to $l=100$, where the comparison between coupled quintessence for different data sets and the $\Lambda$CDM model is presented.
 If the coupling between dark energy and other cosmic components persists into late cosmic times, the LISW effect may be affected. Changes in the late-time acceleration of the universe due to coupled quintessence could leave imprints on the CMB power spectrum at smaller angular scales. The presence of coupled quintessence might influence the behavior of acoustic oscillations in the primordial plasma. This can lead to alterations in the positions and amplitudes of the peaks and troughs in the CMB power spectrum associated with acoustic oscillations. As we can see in Fig 1, for $l \ge 200$, the first peak shifts down with respect to the $\Lambda $CDM model. We can conclude if we consider the models which are scalar fields coupled with neutrinos, create a change in the distribution of matter in the universe. Coupled quintessence can impact the growth of large-scale structures in the universe. This, in turn, affects the distribution of matter at the time of recombination, influencing the CMB temperature fluctuations. Changes in the growth rate of structure may be reflected in the higher multipoles of the CMB power spectrum.	
 Of course, the neutrinos do not scale like a matter, since the pressure does not vanish completely at the beginning of the matter-dominated era and also because of the coupling to the scalar field. Because the abundance of neutrinos in the Universe is very high, when they turn into non-relativistic particles, they can create a gravitational field resulting from their mass density. As a result, it can be concluded that neutrinos play an important role in changing from the radiation-dominated era to the matter-dominated era. Figs. (2-5), shows that $ \Omega_{\nu} $ in redshift interval [180-400] become non-zero. As we can see  in Fig. (6), this interval is located in the region where the universe is entering the era of matter-dominated. It can be concluded that when neutrinos become non-relativistic, considering that they are the most abundant after photons, they can created the gravitational field and play a very effective role in the dominance of matter. During the period of matter dominated, the mass of neutrinos changes between different values. 
 We are interested in investigating the effect of the mass of neutrinos in the recent universe. In the field of cosmology, one can find a connection between bulk flow and coupled quintessence. This connection is mainly related to the study of large-scale structures present in the universe and the dark energy that governs the universe's expansion. The link between coupled quintessence and bulk flow is rooted in the dynamic nature of the dark energy component, which can impact the evolution of large-scale structures. Variations in the universe's rate of expansion and gravitational interactions among cosmic components can imprint discernible patterns on the observed bulk flow. These findings underscore the importance of acknowledging the interplay between dark energy and the evolution of the universe's large-scale structure.	We look at figures 2 to 5 again. In the redshift range of 0.001 to 1.4, we consider two scenarios. In the first scenario, we consider the direction and magnitude of the bulk flow without the coupling of neutrinos with the quintessence field, and in the second scenario, we do this in the presence of the scalar field coupling with neutrinos.

The results obtained in the first scenario are plotted in Fig.(7). In this figure, in the local Universe($0.001<z<0.1$) the direction of bulk flow is in the direction of CMB dipole($ (l,b)=(276,30) $). Also, our results obtained for each redshift are in broad agreement with the results obtained by other scientists. For redshift $ 0.015<z<0.035 $ (\cite{Kashlinsky1},\cite{Barandela}). For redshift $ 0.015<z<0.06 $ (\cite{Feldman},\cite{Nusser},\cite{Macaulay},\cite{Watkins},\cite{Shapley},\cite{Yarahmadi2}).  For redshift $ 0.015<z<0.1 $ (\cite{Wang}) and finally for redshift 
$ 0.015<z<0.1 $ (\cite{Cai}\cite{Mariano}\cite{Chang}\cite{Yang}). But beyond of local universe, the direction of bulk flow is in disagreement with the direction of dark energy dipole. \\
In the second scenario, in low redshifts, the direction of bulk flow is in the direction of the CMB dipole, and also on a large scale, the direction of this flow is in broad agreement with the direction of the dark energy dipole(Fig. 8). As we can see in Table  4, the velocity of bulk flow increases from redshift 0.001 to 0.1, and from 0.001 to 1.4, this velocity value decreases. To achieve more accurate results and eliminate the effect of lower redshifts, we use the tomography method. In this method, we select the different redshift shells which are not related to each other. We consider the two regions: 1-  $ z < 0.1 $, 2- $ 0.1<z<1.4 $. Figs. (9,10) show the direction and amplitude of bulk velocity in $ 0.025<z<0.045 $, $ 0.045<z<0.06 $, and $ 0.06<z<0.1 $. As the neutrinos start to become non-relativistic, part of the energy of the neutrinos is transferred to kinetic energy for the scalar field. After that, neutrinos become non-relativistic at about the same temperature as the neutrino mass and scale similarly to dark energy, but differently compared to dark matter. Here the kinetic energy dominates the dynamics of the scalar field. While the other energy densities decay away, dark energy density begins to dominate.
As can be seen in Figure 6, at redshift z=1, the Universe starts entering the dark energy-dominated era.  By closely analyzing the ratio of neutrino density to redshift in Figures 2 to 5, it can be concluded that as the density of neutrinos decreases, their gravitational field becomes weaker, and as a result, the velocity of the bulk flow increases in redshift $ 0.1 <z< 1 $. Figures 11 and 12 confirm this conclusion well.

Also, at the redshift $ 1<z< 1.4 $, neutrino density decreases and the amplitude of bulk velocity increases. Figure 17 demonstrate the bulk flow as a fraction of redshift from likelihood analysis fo $ z>0.1 $(The Large - scale structure). In this figure we can see the changes in amplitude of  bulk flow magnitude  $ 0.1<z< 1.4 $.
 Also, the bulk flow direction is in good agreement with the direction of the dark energy dipole in this redshift. According to Figures 7 and 8, the difference between the direction of the bulk flow at the redshift is less than 0.1 in the case where the quintessence field is coupled with neutrinos and the uncoupled state is small, but as we go to higher redshifts, this difference increases.  At redshift 1.4, this difference is very large, and as we can see, in the coupled model, the obtained direction is very close to the dark energy dipole direction. All results are shown in table 4.
	
	\begin{figure}
		\centering
		\includegraphics[width=8 cm]{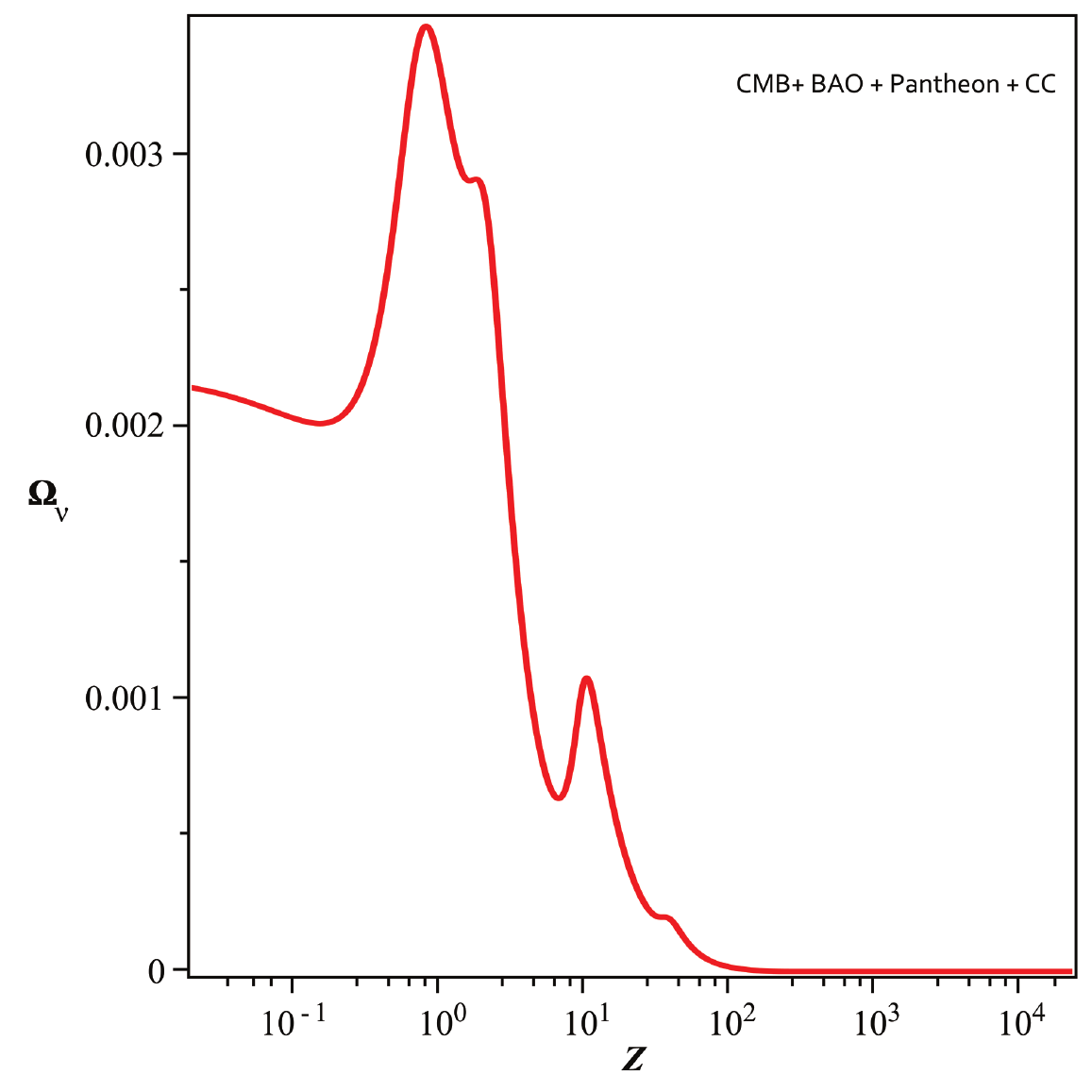}
		\vspace{-0.12cm}
		\caption{\small{The evolution of the  neutrino density respect to redshift z for CMB + BAO + CC + Pantheon + Anlysis}}\label{fig:omegam2}
	\end{figure}

	\begin{figure}
		\centering
		\includegraphics[width=8 cm]{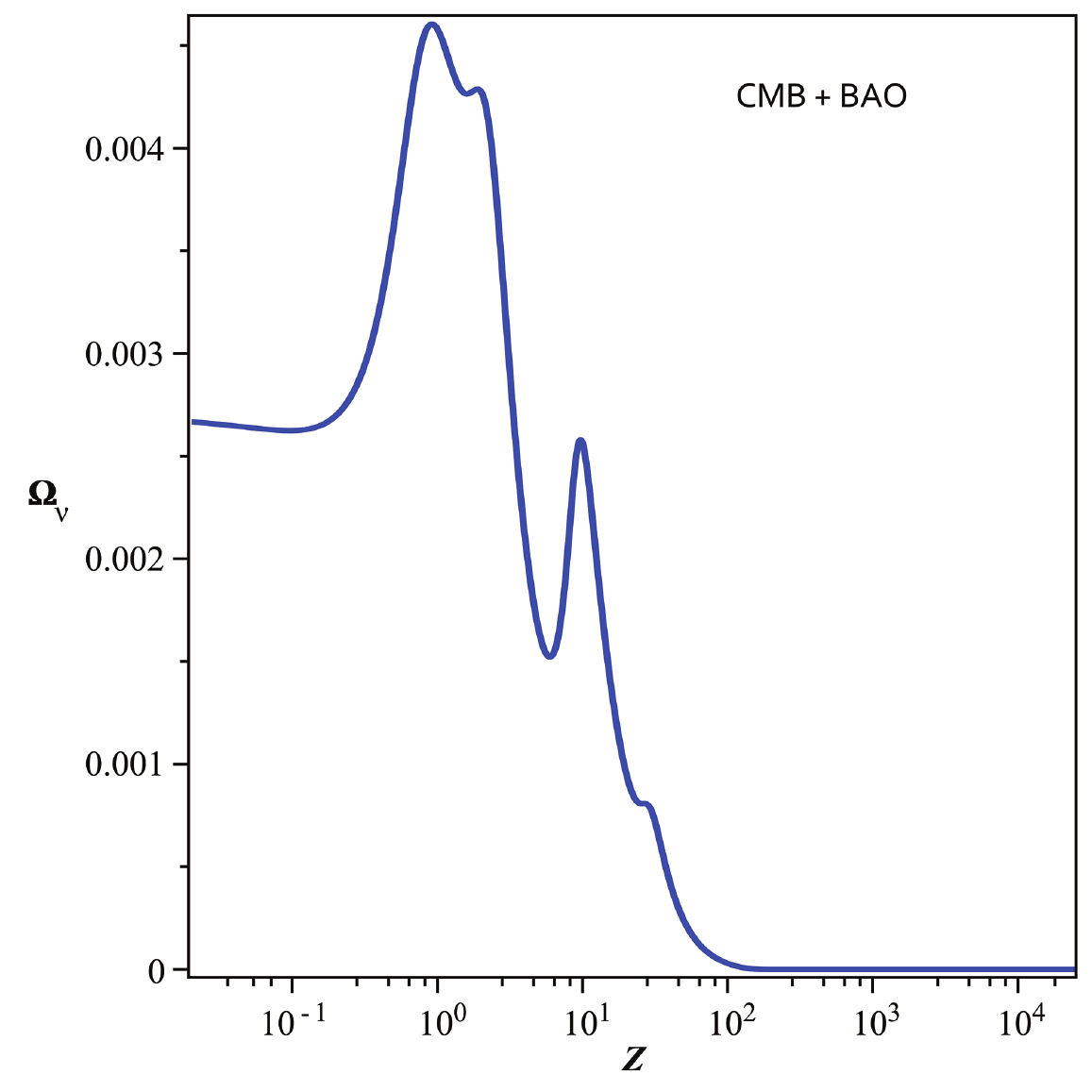}
		\vspace{-0.12cm}
		\caption{\small{The evolution of the  neutrino density respect to redshift z for CMB + BAO }}\label{fig:omegam2}
	\end{figure}

	\begin{figure}
		\centering
		\includegraphics[width=8 cm]{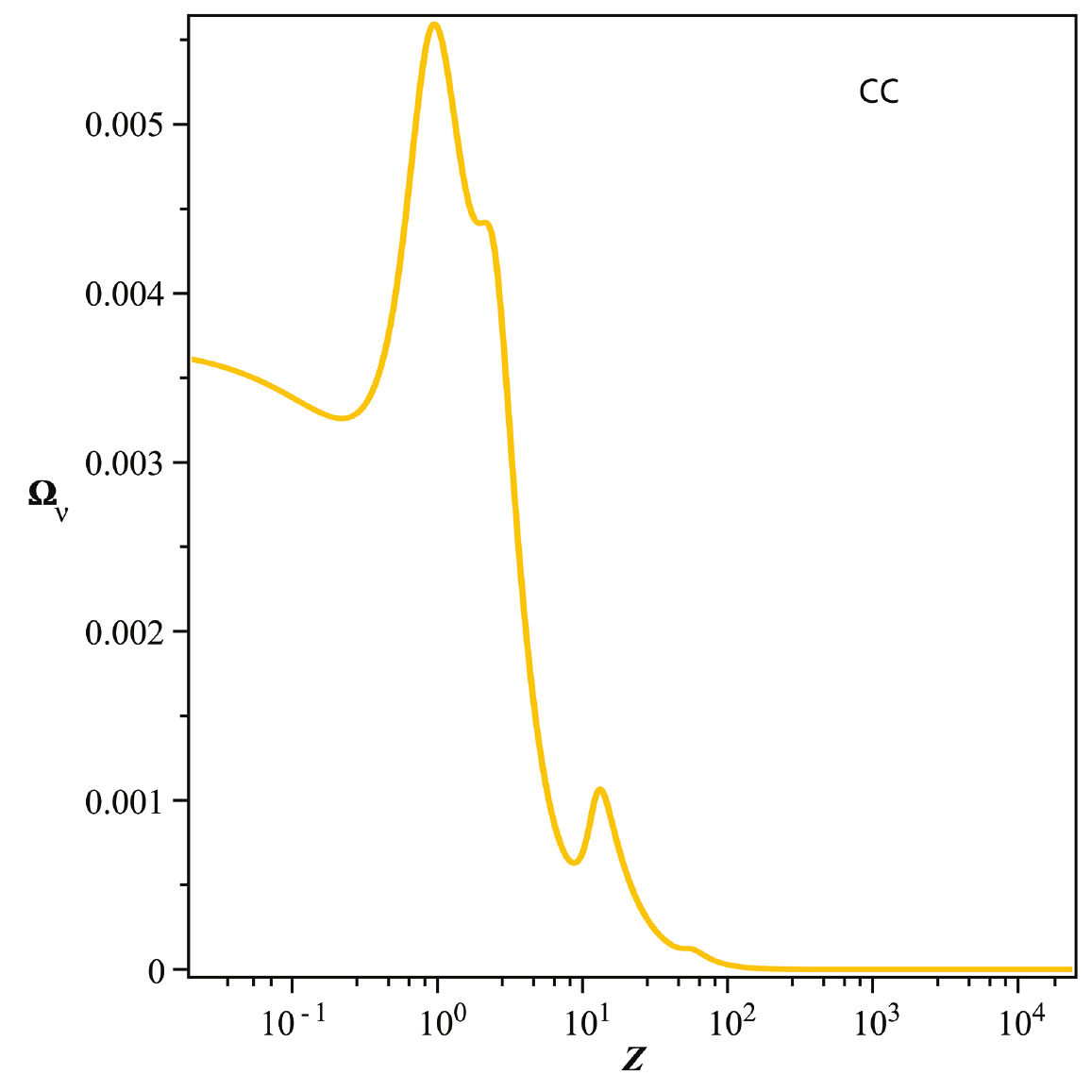}
		\vspace{-0.12cm}
		\caption{\small{The evolution of the  neutrino density respect to redshift z for CC data}}\label{fig:omegam2}
	\end{figure}

	\begin{figure}
		\centering
		\includegraphics[width=8 cm]{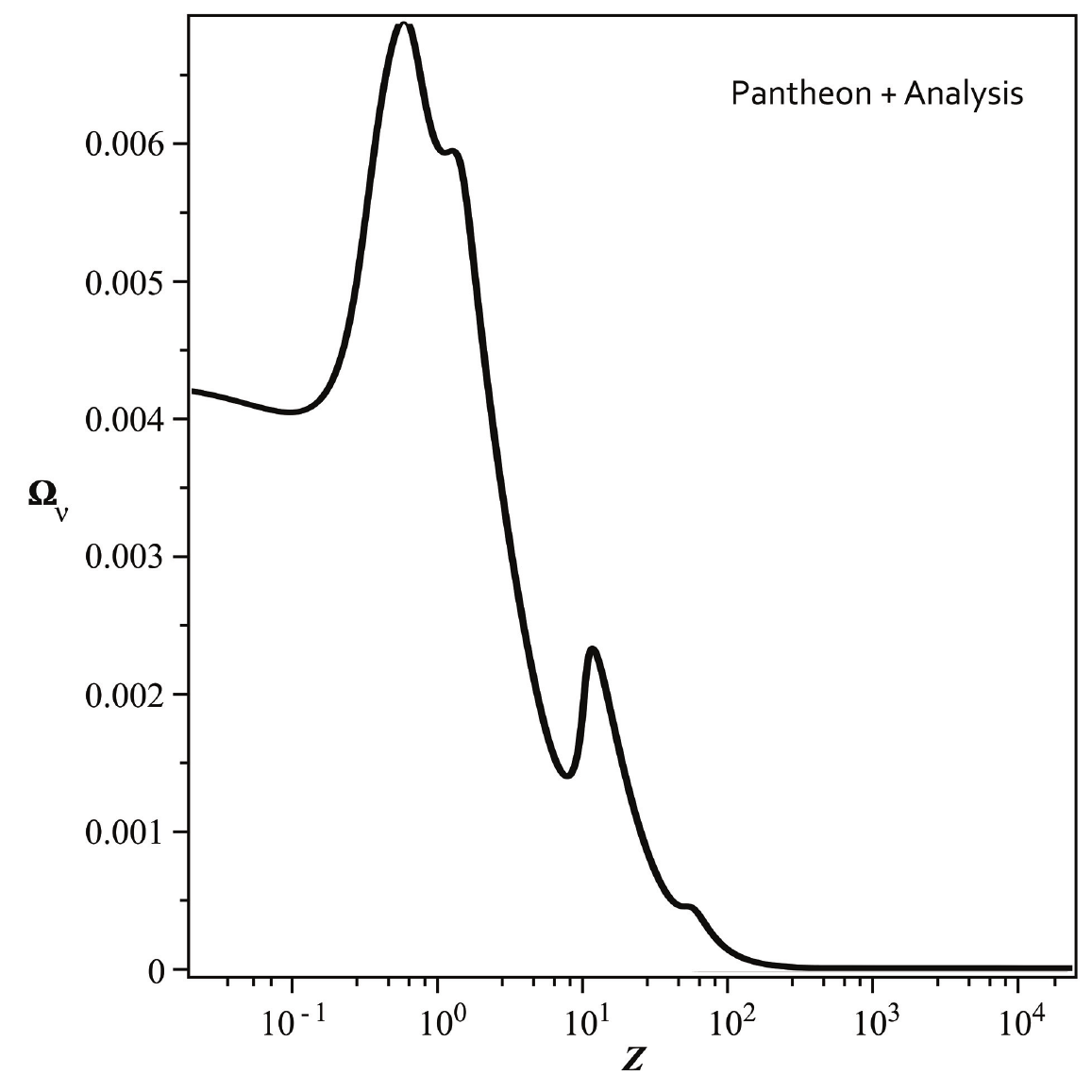}
		\vspace{-0.12cm}
		\caption{\small{The evolution of the  neutrino density respect to redshift z for Pantheon + Anlysis}}\label{fig:omegam2}
	\end{figure}

	\begin{figure*}
		\centering
		\includegraphics[width=13 cm]{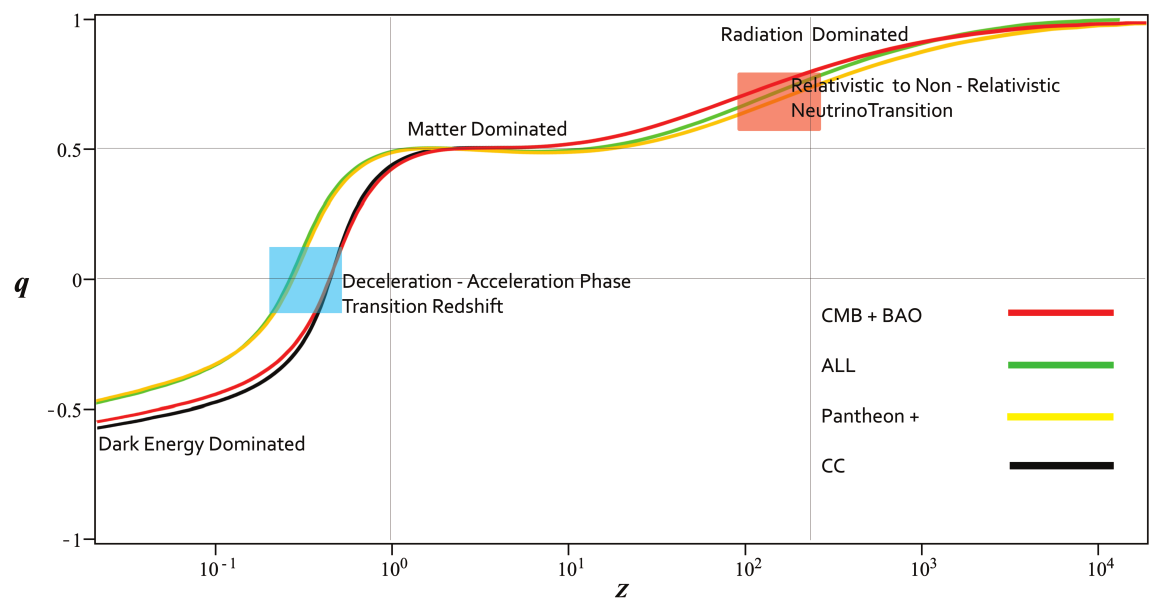}
		\vspace{-0.12cm}
		\caption{\small{The evolution deceleration parameter  respect to redshift z for different dataset}}\label{fig:omegam2}
	\end{figure*}

	\begin{table*}
		\scriptsize
		\caption{Results from the  analysis 
			between the models (Q: Quintessence and CQ: Coupled Quintessence with neutrinos), for different redshift shells  }  
		\centering 
		\begin{tabular}{|c@{\hspace{6mm}}|c@{\hspace{6mm}}|c@{\hspace{6mm}}|c@{\hspace{6mm}}|c@{\hspace{6mm}}|c@{\hspace{6mm}}
				|c@{\hspace{6mm}}|c@{\hspace{6mm}}|c@{\hspace{6mm}}|c@{\hspace{6mm}}|c@{\hspace{6mm}}|c@{\hspace{6mm}}|c@{\hspace{6mm}}
				|c@{\hspace{6mm}}|c@{\hspace{6mm}}|c@{\hspace{6mm}}|c@{\hspace{6mm}}|c@{\hspace{6mm}}|c@{\hspace{6mm}}|c@{\hspace{6mm}}
				|c@{\hspace{6mm}}|c@{\hspace{6mm}}|c@{\hspace{6mm}}|c@{\hspace{6mm}}|c@{\hspace{6mm}}|c@{\hspace{6mm}}|c@{\hspace{6mm}}
				|c@{\hspace{6mm}}|c@{\hspace{6mm}}|c@{\hspace{6mm}}|c@{\hspace{6mm}}|c@{\hspace{6mm}}|c@{\hspace{6mm}}|c@{\hspace{6mm}}
				|c@{\hspace{6mm}}|c@{\hspace{6mm}}|c@{\hspace{6mm}}|c@{\hspace{6mm}}|c@{\hspace{6mm}}|c@{\hspace{6mm}}|c@{\hspace{6mm}}
				|c@{\hspace{6mm}}|c@{\hspace{6mm}}|c@{\hspace{6mm}}|c@{\hspace{6mm}}} 
			\hline\hline 
			$Range$ &$Model$   & $l^{o}$ \ &  $b^{o}$ \ &  $h_{0}$ \ & $v_{bulk}$ \   \\
			\rule{0pt}{4mm}%
			$$ & $$  & $$ \ &  $$ \ & $km s^{-1}Mpc^{-1}$ \ & $km s^{-1}$ \  \\
			\hline 
			
			\hline 
			$0.001<z<0.035$  & \\
			\hline 
			& Q & $267^{+20}_{-20}$ &$21^{+16}_{-16}$ & $0.7\pm0.12$ & $170^{+32}_{-28}$ \\
			\rule{0pt}{4mm}%
			& CQ &$292^{+22}_{-22}$ &$15.5^{+17}_{-17}$ &  $0.715\pm0.12$ & $258^{+82}_{-70}$  \\
			\hline 
			$0.001<z<0.06$ & \\
			\hline 
			& Q & $300^{+18}_{-18}$ &$14^{+13}_{-13}$&  $0.701\pm0.17$ & $177^{+33}_{-30}$    \\
			\rule{0pt}{4mm}%
			& CQ  & $274^{+15}_{-15}$ &$26^{+11}_{-11}$& $0.702\pm0.16$ & $267^{+63}_{-37}$   \\
			\hline 
			$0.001<z<0.1$  &\\
			\hline 
			& Q & $ 302^{+20}_{-20}$ &$29^{+10}_{-10}$ &  $0.697\pm0.2$ &  $225^{+25}_{-20}$      \\
			\rule{0pt}{4mm}%
			& CQ  & $ 272^{+17}_{-17}$ &$33^{+12}_{-12}$&  $0.714\pm0.14$ &  $264^{+64}_{-46}$   \\
			\hline 
			$0.001<z<1.4$  & \\
			\hline 
			& Q & $ 268^{+12}_{-12}$ &$45^{+10}_{-10}$ & $0.7\pm0.141$ &  $265^{+45}_{-22}$   \\
			\rule{0pt}{4mm}%
			& CQ  &$ 308^{+10}_{-30}$ &$-14^{+7}_{-7}$ &  $0.716\pm0.102$ &  $253^{+67}_{-63}$   \\
			\hline 
			$0.025<z<0.045$ & \\
			\hline 
			& Q & $324^{+15}_{-18}$ &$40^{+12}_{-11}$ &  $0.706\pm0.19$ & $180^{+25}_{-30}$  \\
			\rule{0pt}{4mm}%
			& CQ& $235^{+17}_{-14}$ &$25^{+13}_{-10}$ &  $0.725\pm0.26$ & $265^{+86}_{-86}$  \\
			\hline 
			$0.045<z<0.06$  & \\
			\hline 
			& Q & $ 350^{+17}_{-15}$ &$10^{+13}_{-11}$ & $0.7\pm0.127$ &  $340^{+110}_{-140}$   \\
			\rule{0pt}{4mm}%
			& CQ  &$272^{+12}_{-18}$ &$30^{+22}_{-20}$ & $0.702\pm0.113$ &  $420^{+122}_{-119}$  \\
			\hline 
			$0.06<z<0.1$ &  \\
			\hline 
			& Q & $ 128^{+14}_{-14}$ &$7^{+11}_{-11}$ &  $0.698\pm0.18$ &  $620^{+80}_{-80}$  \\
			\rule{0pt}{4mm}%
			& CQ &$ 282.5^{+15}_{-14}$ &$27^{+15}_{-13}$ &  $0.701\pm0.14$ &  $719^{+111}_{-119}$ \\
			\hline 
			$0.1<z<0.2$ & \\
			\hline 
			& Q & $ 318^{+10}_{-10}$ &$-15^{+16}_{-14}$ &  $0.7\pm0.17$ &  $708^{+108}_{-92}$ \\
			\rule{0pt}{4mm}%
			& CQ &$ 265^{+16}_{-14}$ &$29^{+7}_{-10}$ & $0.713\pm0.152$ &  $982^{+86}_{-114}$  \\
			\hline 
			$0.2<z<0.4$ & \\
			\hline 
			& Q & $310^{+23}_{-19}$ &$12^{+10}_{-10}$ &  $0.699\pm0.133$ & $750^{+95}_{-98}$   \\
			\rule{0pt}{4mm}%
			& CQ & $298^{+14}_{-14}$ &$13^{+9}_{-8}$ &  $0.722\pm0.149$ & $995^{+215}_{-270}$  \\
			\hline 
			$0.4<z<0.6$  & \\
			\hline 
			& Q & $337^{+22}_{-20}$ &$9^{+11}_{-11}$ &  $0.703\pm0.14$ & $740^{+111}_{-117}$  \\
			\rule{0pt}{4mm}%
			& CQ  &$358^{+29}_{-25}$ &$12^{+17}_{-14}$ &  $0.72\pm0.196$ & $2350^{+196}_{-189}$  \\
			\hline 
			$0.6<z<1$ & \\
			\hline 
			& Q & $294^{+19}_{-19}$ &$12^{+10}_{-10}$ &  $0.701\pm0.173$ & $842^{+109}_{-114}$  \\
			\rule{0pt}{4mm}%
			& CQ &$302.5^{+23}_{-19}$ &$28^{+7}_{-12}$ &  $0.714\pm0.141$ & $2460^{+185}_{-173}$  \\
			\hline 
			$1<z<1.4$  &\\
			\hline 
			& Q & $ 119^{+18}_{-18}$ &$46^{+15}_{-15}$ & $0.705\pm0.176$ &  $900^{+121}_{-120}$    \\
			\rule{0pt}{4mm}%
			& CQ  &$ 301^{+15}_{-11}$ &$-7^{+8}_{-10}$ & $0.711\pm0.164$ &  $2280^{+106}_{-110}$    \\
			\hline 
			\hline 
		\end{tabular}\\
		\label{table:2} 
	\end{table*}

	\begin{figure}
		\centering
		\includegraphics[width=12 cm]{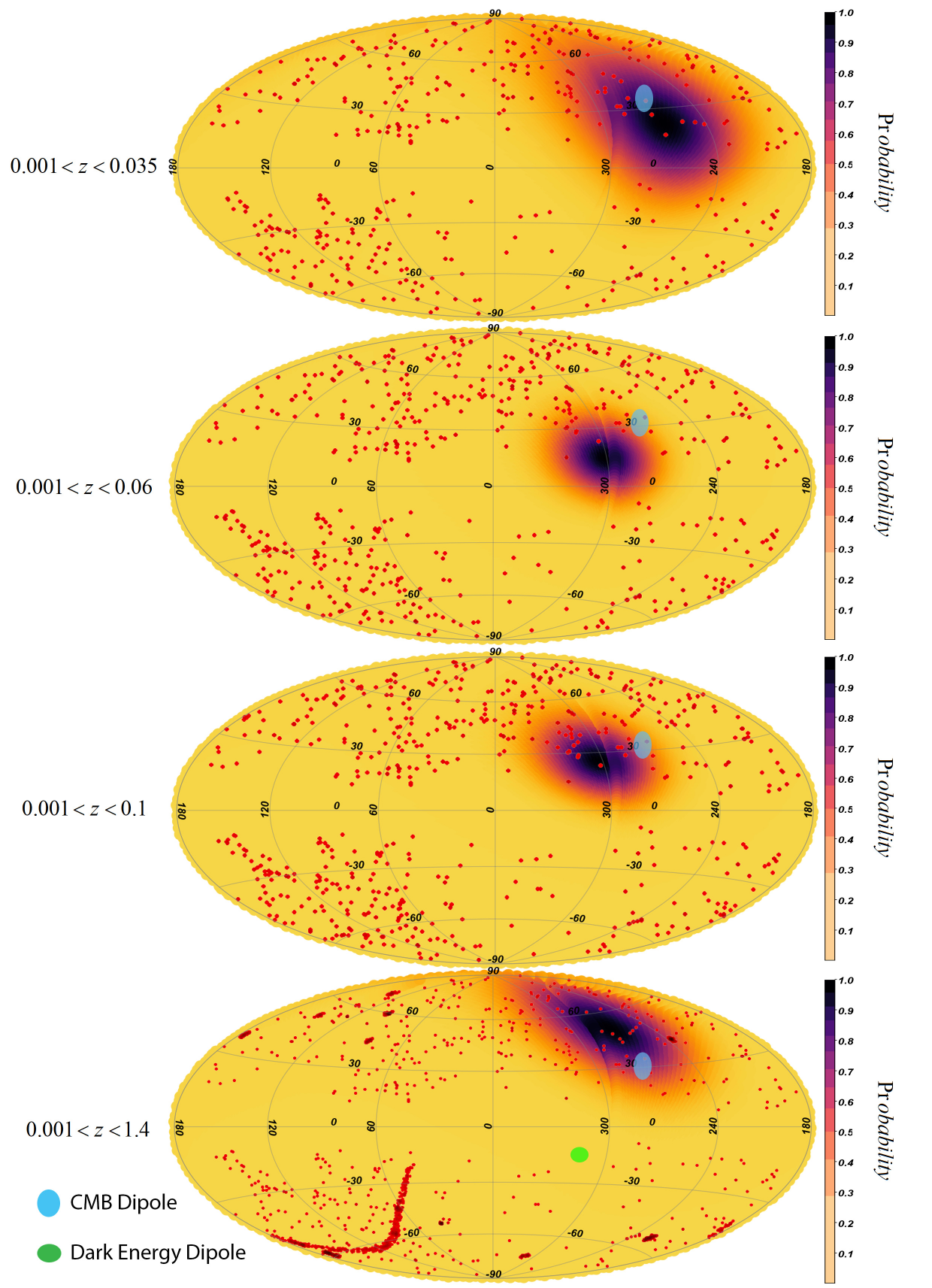}
		\vspace{-0.12cm}
		\caption{\small{The direction of the bulk flow for redshifts $ 0.001<z<0.035 $, $ 0.001<z<0.06 $, $ 0.001<z<0.1 $, $ 0.001<z<1.4 $ for quintessence model without neutrinos. The red circle denotes supernova type Ia. }}\label{fig:omegam2}
	\end{figure}
	\begin{figure}
		\centering
		\includegraphics[width=12 cm]{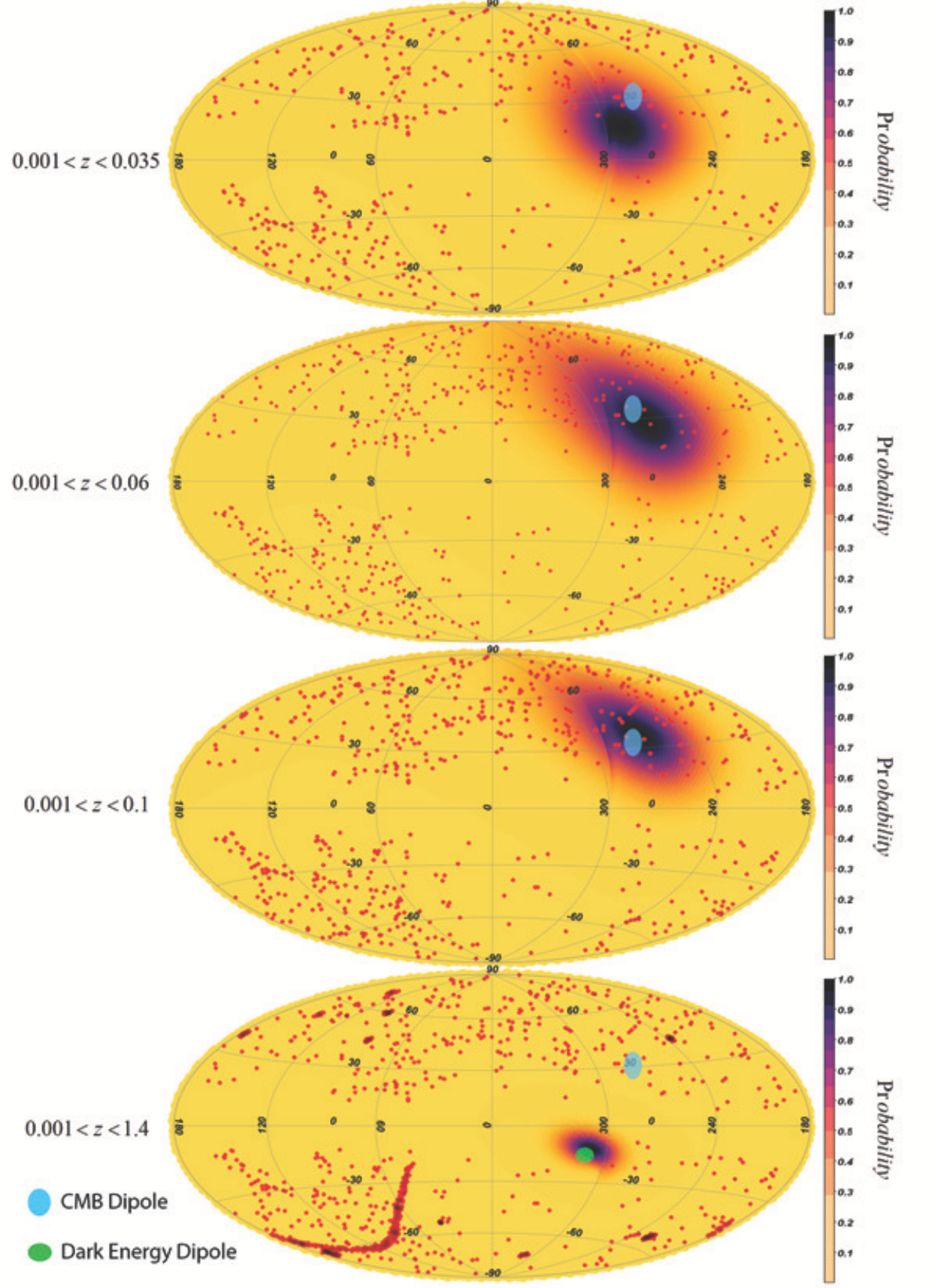}
		\vspace{-0.12cm}
		\caption{\small{The direction of the bulk flow for redshifts $ 0.001<z<0.035 $, $ 0.001<z<0.06 $, $ 0.001<z<0.1 $, $ 0.001<z<1.4 $ for quintessence coupled with neutrinos. The red circle denotes supernova type Ia.}}\label{fig:omegam2}
	\end{figure}

	\begin{figure}
		\centering
		\includegraphics[width=11 cm]{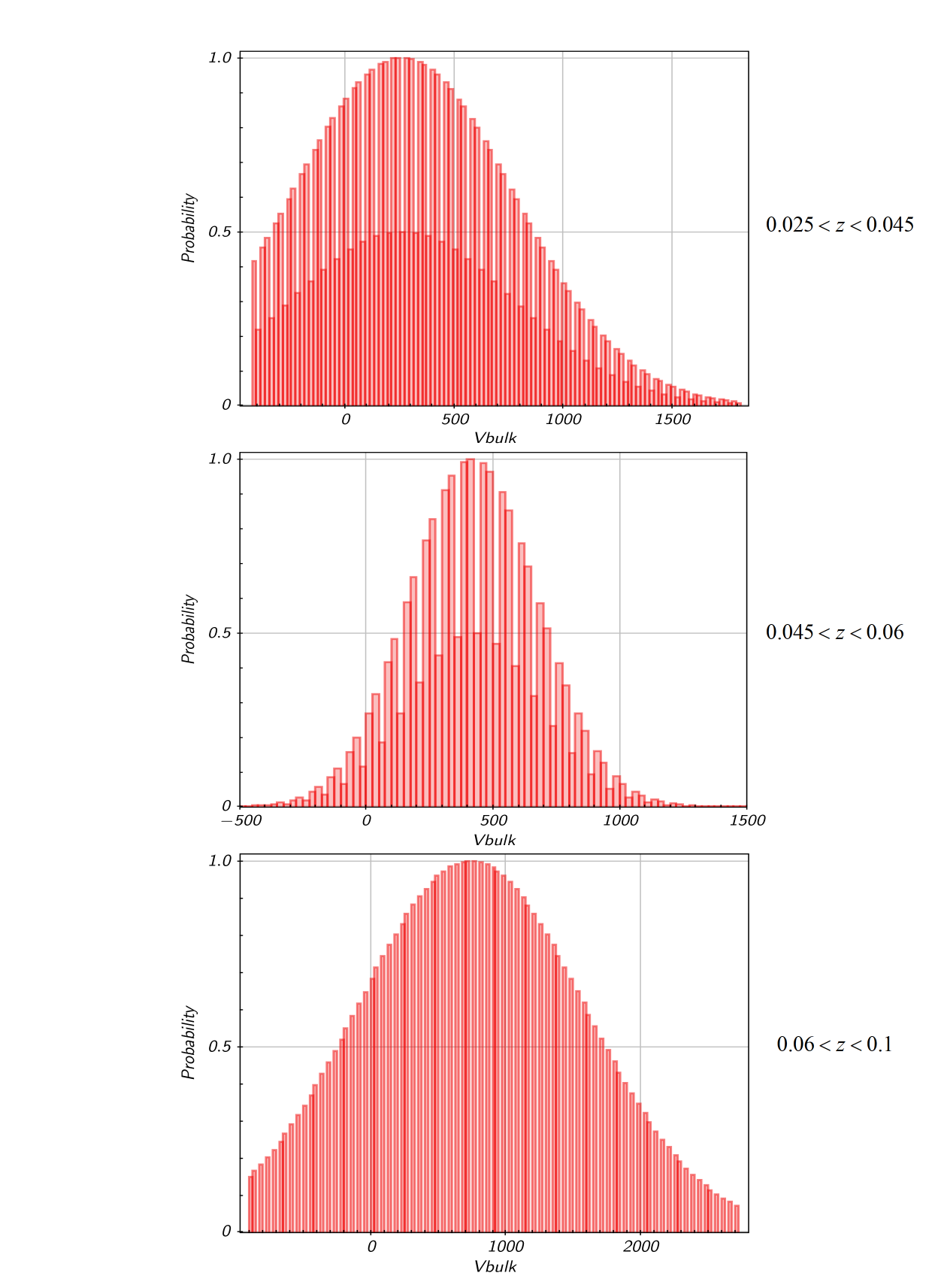}
		\vspace{-0.12cm}
		\caption{\small{The amplitude of the bulk flow velocity for $ z < 0.1 $ }}\label{fig:omegam2}
	\end{figure}

	\begin{figure}
		\centering
		\includegraphics[width=13 cm]{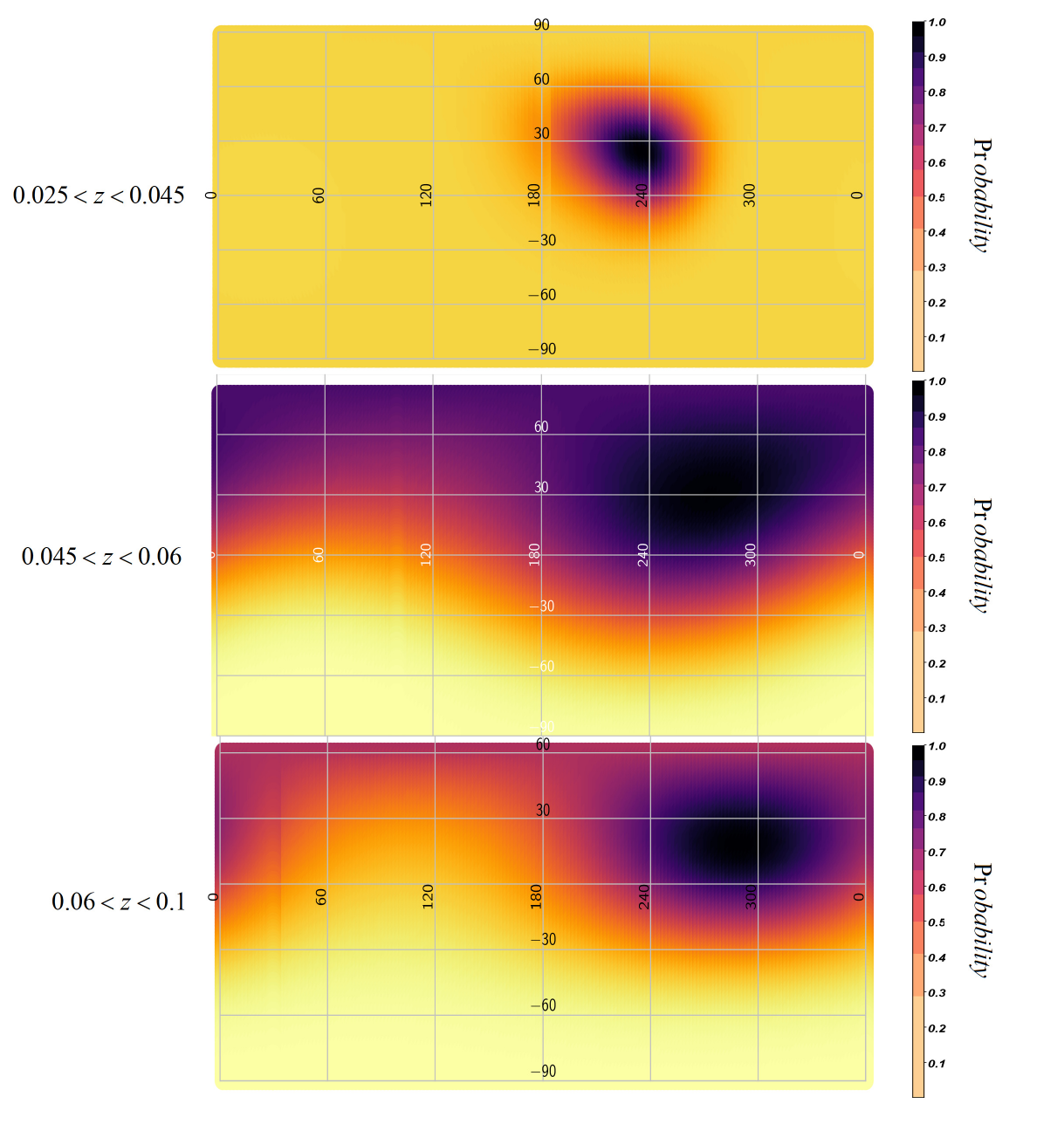}
		\vspace{-0.12cm}
		\caption{\small{The direction of  bulk flow  for $ z < 0.1 $}}\label{fig:omegam2}
	\end{figure}

	\begin{figure}
		\centering
		\includegraphics[width=12 cm]{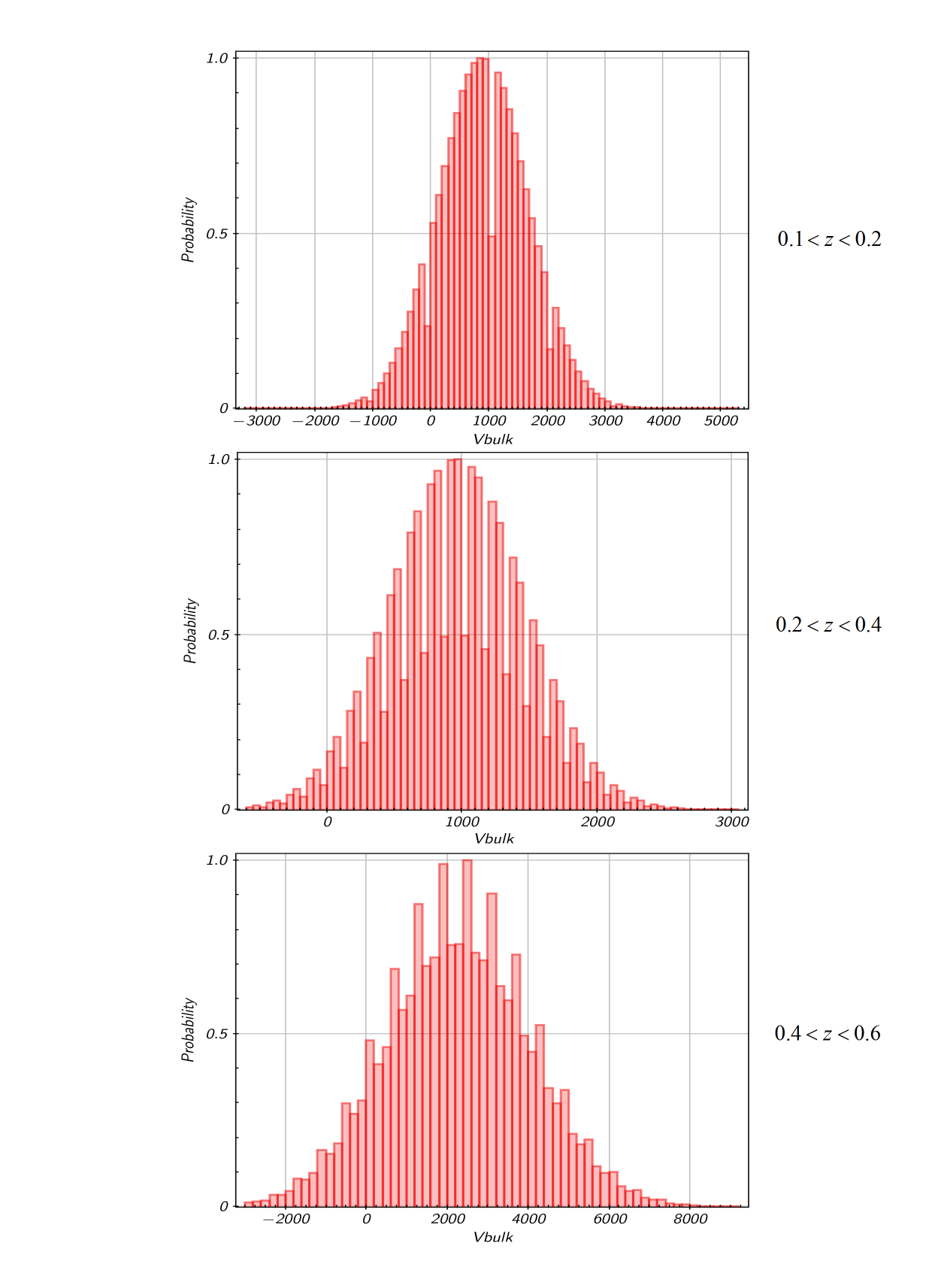}
		\vspace{-0.12cm}
		\caption{\small{The amplitude of the bulk flow velocity for $ z > 0.1 $}}\label{fig:omegam2}
	\end{figure}
	
	\begin{figure}
		\centering
		\includegraphics[width=12 cm]{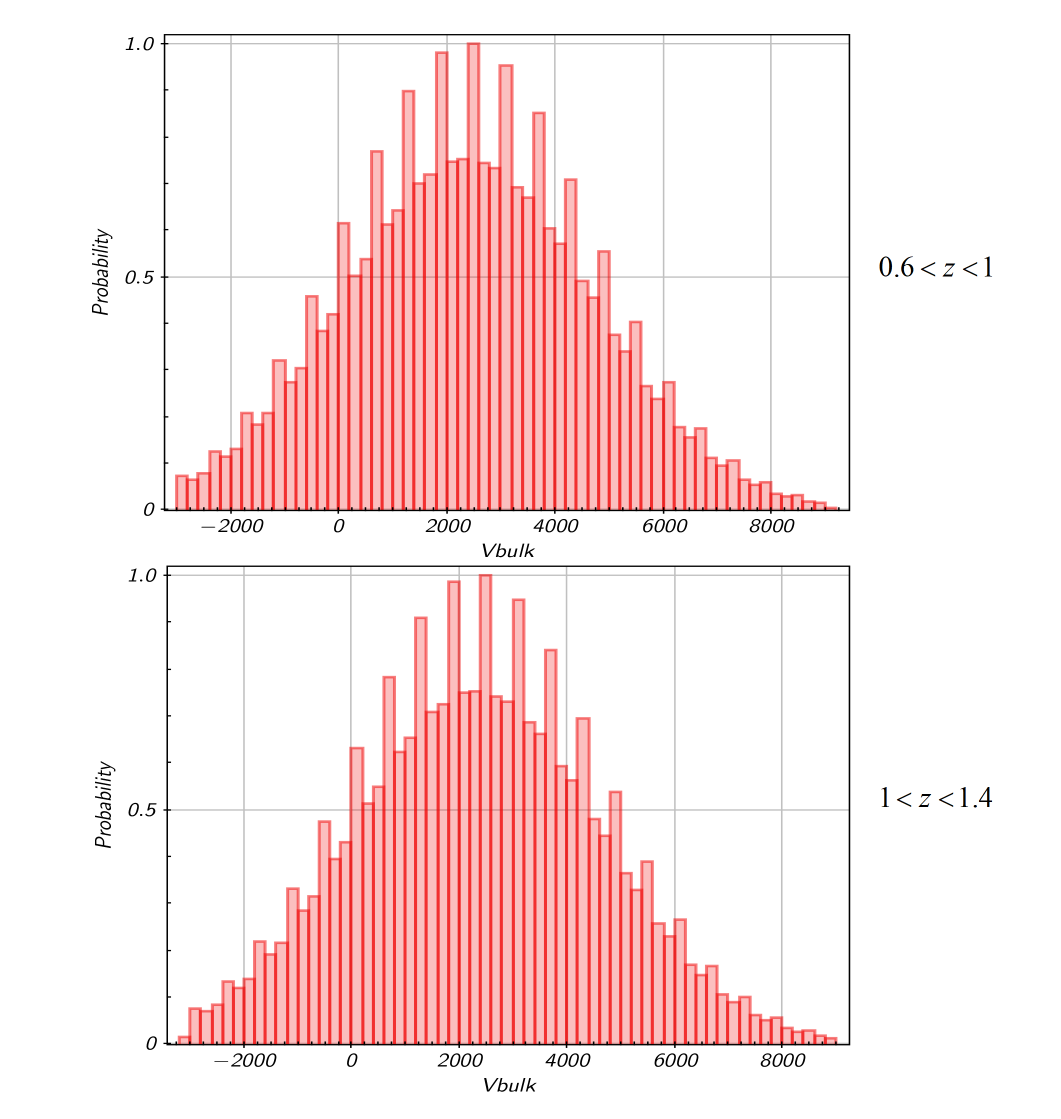}
		\vspace{-0.12cm}
		\caption{\small{The amplitude of the bulk flow velocity for $ z > 0.1 $}}\label{fig:omegam2}
	\end{figure}
	
	\begin{figure}
		\centering
		\includegraphics[width=13 cm]{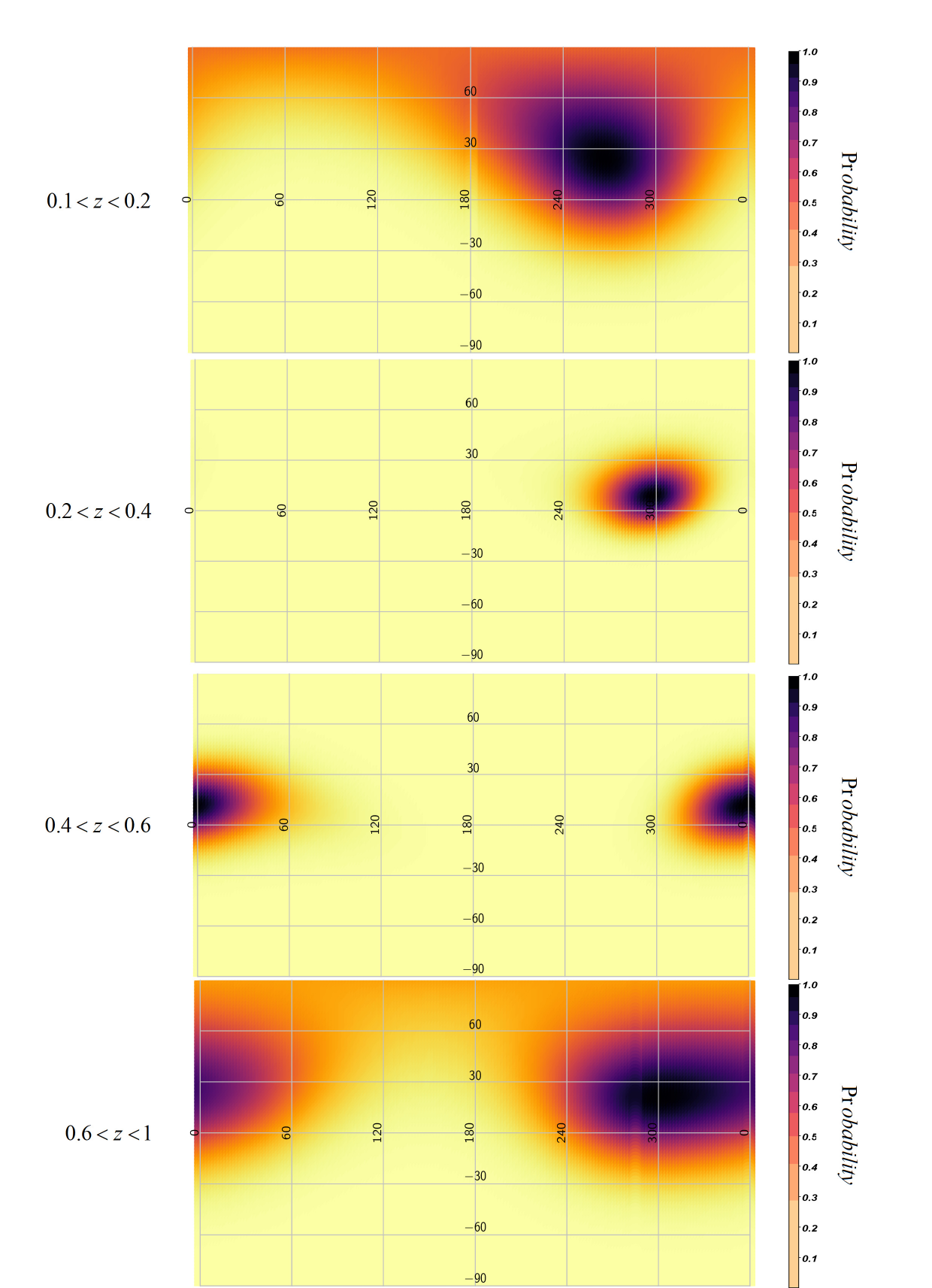}
		\vspace{-0.12cm}
		\caption{\small{The direction of  bulk flow  for $ z > 0.1 $}}\label{fig:omegam2}
	\end{figure}

	\begin{figure}
		\centering
		\includegraphics[width=13 cm]{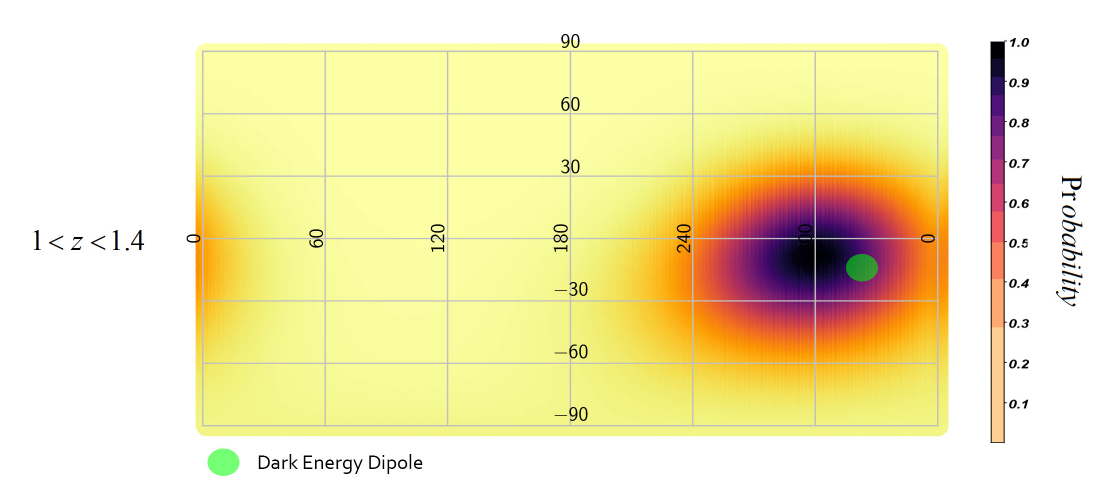}
		\vspace{-0.12cm}
		\caption{\small{The direction of  bulk flow  for $ z > 0.1 $}}\label{fig:omegam2}
	\end{figure}

		\begin{figure}
		\centering
		\includegraphics[width=10 cm]{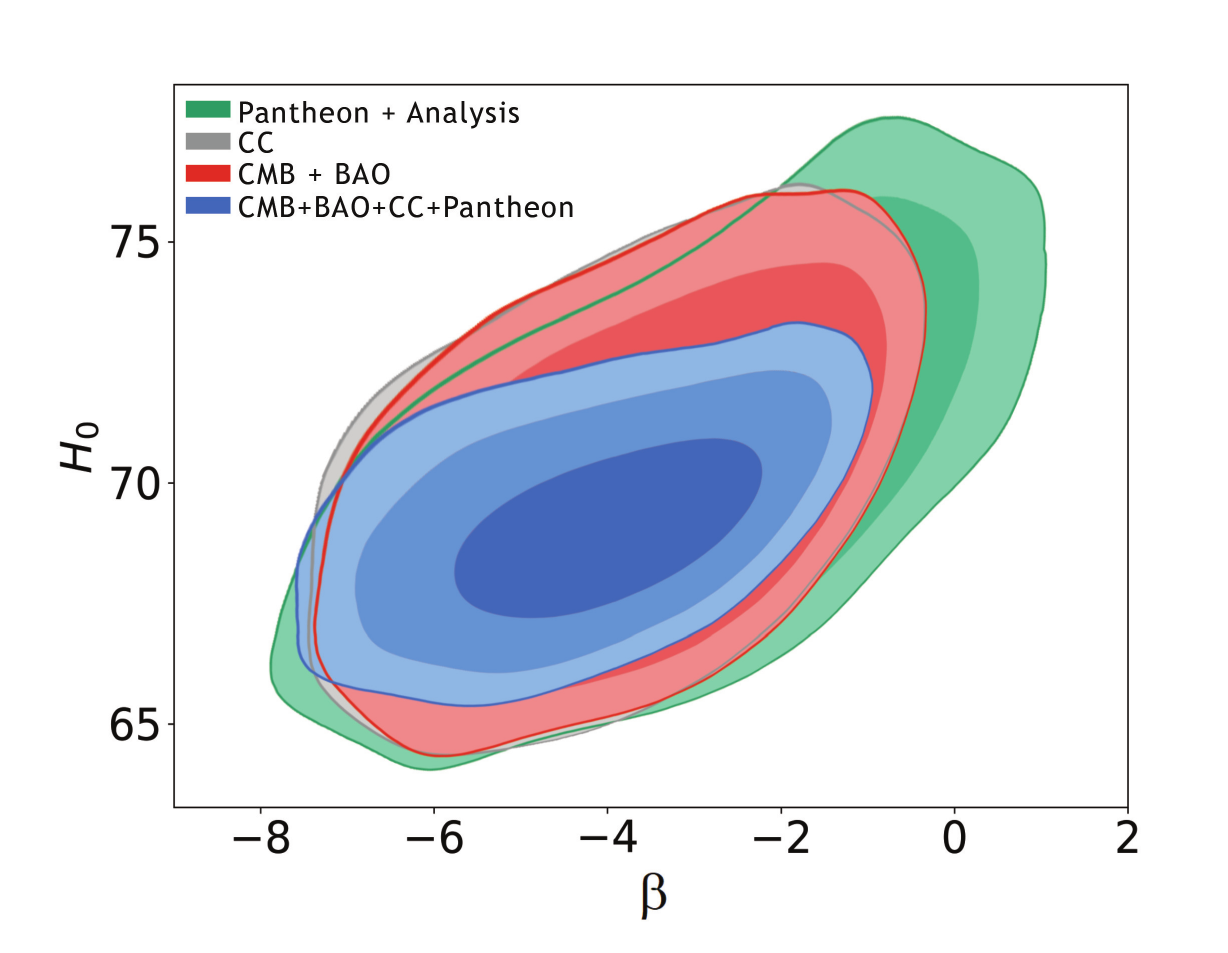}
		\vspace{-0.12cm}
		\caption{\small{Comparison between $\beta$ value in different combination for coupled quintessence with neutrinos}}\label{fig:omegam2}
	\end{figure}

	\begin{figure}
	\centering
	\includegraphics[width=9.5 cm]{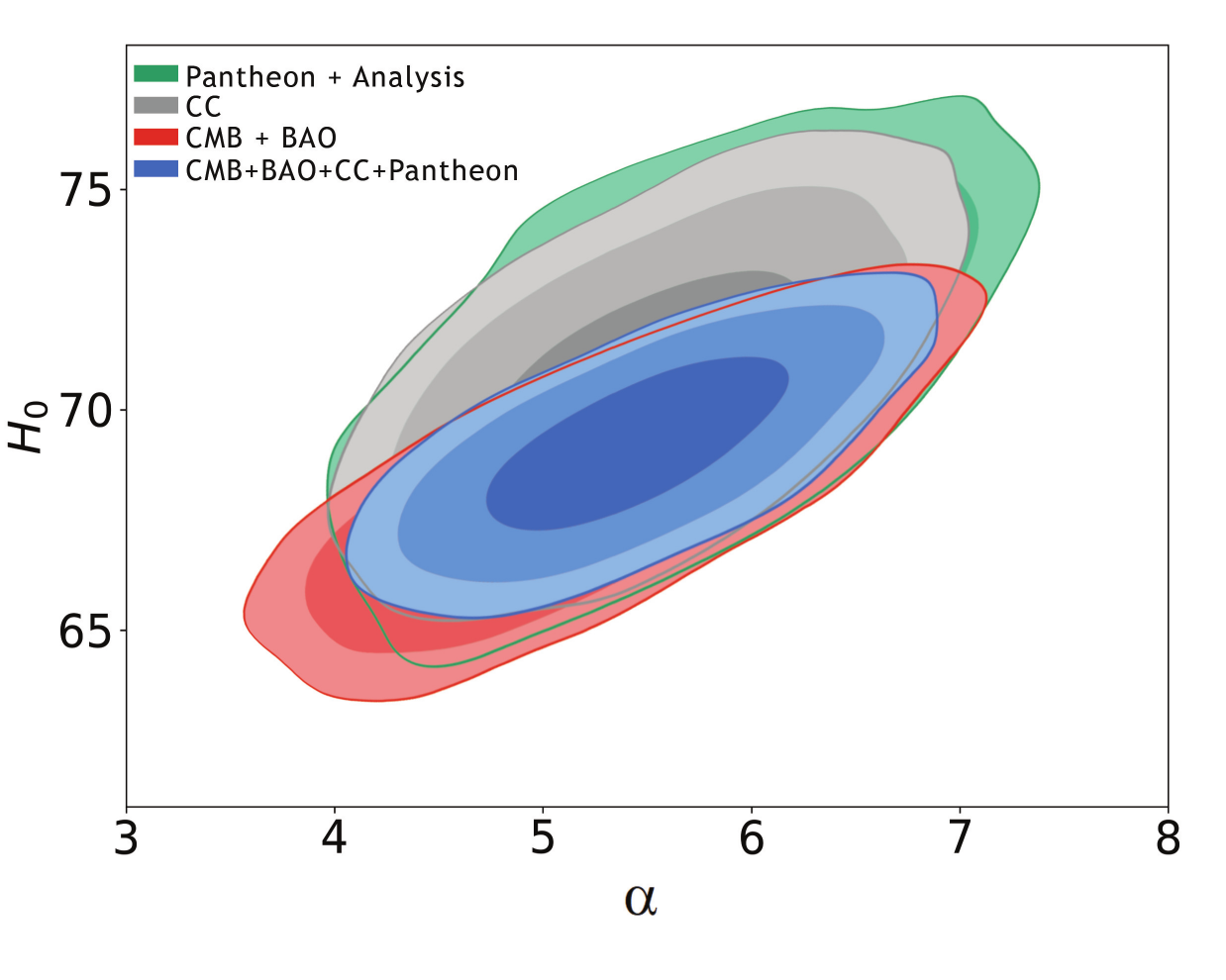}
	\vspace{-0.12cm}
	\caption{\small{Comparison between $\alpha$ value in different combination for coupled quintessence with neutrinos}}\label{fig:omegam2}
\end{figure}

\begin{figure}
	\centering
	\includegraphics[width=10.5 cm]{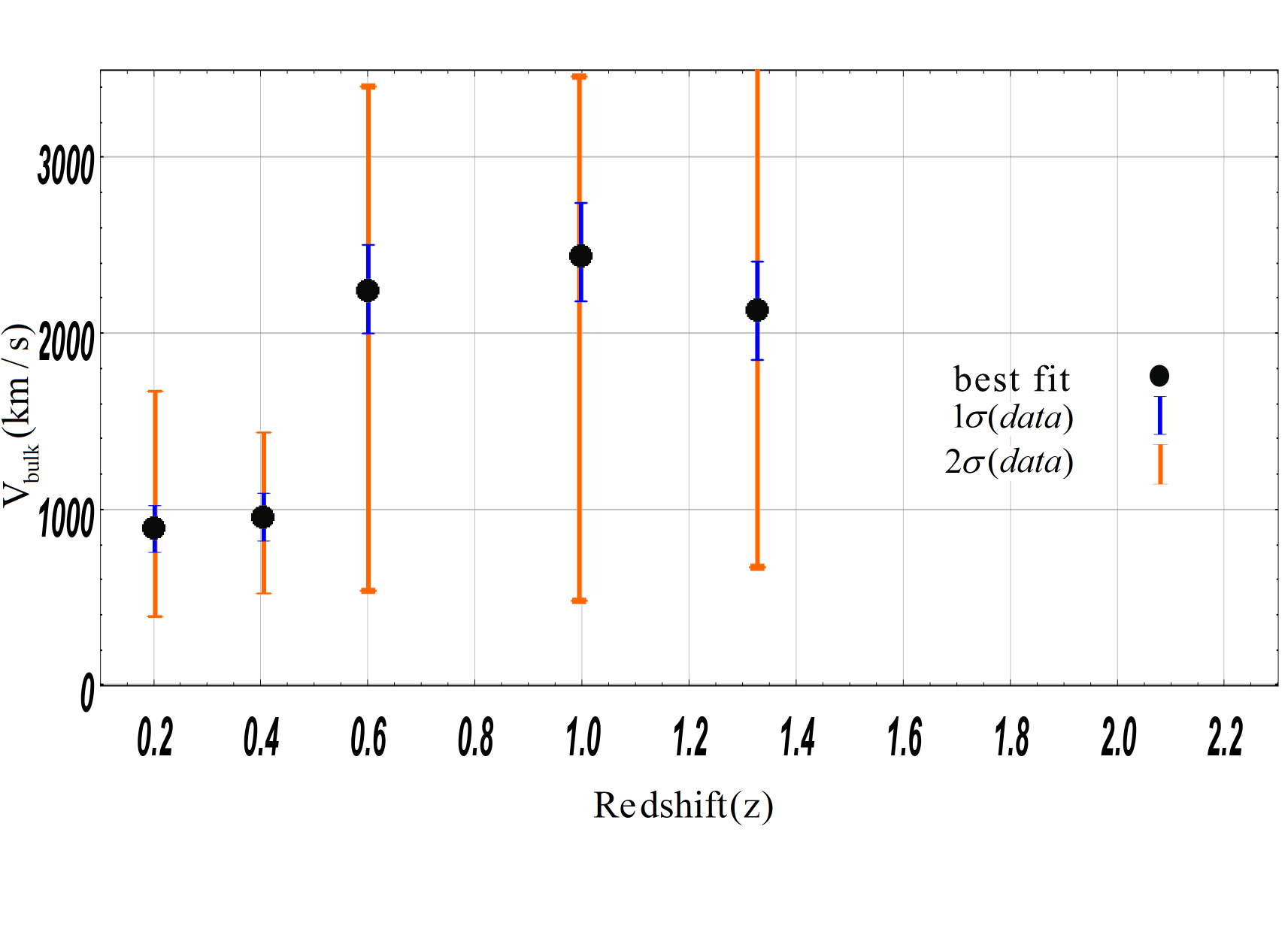}
	\vspace{-0.42cm}
	\caption{\small{The bulk flow as a fraction of redshift from likelihood analysis fo $ z>0.1 $(The Large - scale structure)}}\label{fig:omegam2}
\end{figure}
	
	\section{Conclusion}
Throughout the cosmic timeline, neutrinos originating from the early Universe transitioned from a relativistic phase during its nascent stages to adopting characteristics akin to particles with mass in later epochs. The mass of these neutrinos exerts a discernible influence on the universe's expansion history. In this article, we embark on an exploration of cosmic evolution, delving into the epoch spanning from radiation dominance to the era characterized by the dominance of dark energy, employing the paradigm of neutrino coupling with a scalar field as our investigative tools.

In the first step, constraints on the total mass of neutrinos were derived by coupling the neutrino with a scalar field, yielding the following results:

From the analysis of the CMB + BAO data, we find that $\sum m_{\nu} < 0.115$ eV (95$\%$ CL). Using Pantheon+, we find $\sum m_{\nu} < 0.201$ eV (95$\%$ CL). For CC data, $\sum m_{\nu} < 0.16$ eV (95$\%$ CL), and for the combination of full data (Pantheon+CMB+BAO+CC), we find $\sum m_{\nu} < 0.101$ eV (95$\%$ CL).

The values obtained for ${z_{\rm nr}}$ for different datasets are as follows:
- For the combination data, ${z_{\rm nr}} = 180$
- For Pantheon+ data, ${z_{\rm nr}} = 370$
- For CMB + BAO, ${z_{\rm nr}} = 210$
- For CC data, ${z_{\rm nr}} = 302$

These values are within the region where the Universe enters the matter-dominated era. It was concluded that the neutrinos, upon gaining mass, contributed to the creation of the gravitational field, potentially influencing the observed cosmic evolution.

Moreover, we  focused to parameter $\alpha$, the following constraints have been derived: $\alpha = 5.987^{+1.68}_{-1.27}, \alpha = 5.64^{+1.6}_{-1.3}, \alpha = 5.27^{+1.36}_{-1.23}, \alpha = 5.44^{+1.2}_{-1.1}$, at $68\%$ CL for Pantheon+, CC, CMB + BAO, and Pantheon + CC + CMB + BAO, respectively.
To evoke growing neutrino quintessence, certain conditions must be satisfied:
\begin{itemize}
	\item $V(\sigma)$ must exhibit a negative gradient, inducing an increase in the scalar field value over time. This gradient must be steep enough for $\phi$ to attain sufficiently large values in the late Universe, functioning as dark energy.
	\item $|\alpha|$ must be adequately large when neutrinos transition to a non-relativistic state, enabling $\beta(\rho_\nu - 3p_\nu)$ to act as a robust restoring force, halting the evolution of $\sigma$ as per Eq. (2.18).
\end{itemize}

The best-fitted values of $(\alpha, \lambda)$ adhere to the aforementioned conditions. Despite the seemingly modest value of $\nu$, the role of neutrinos remains crucial in shaping the cosmos' evolution, attributed to their substantial coupling represented by $\alpha$. We listed another bestfitted cosmological parameter in table 3.

Moving forward, the investigation focused on the effect of coupled neutrinos on the late-time evolution of the Universe. Two scenarios were considered in the redshift range of 0.001 to 1.4. In the first scenario, the direction and magnitude of the bulk flow were considered without the coupling of neutrinos with the quintessence field. In the second scenario, this coupling was taken into account.

Results from the first scenario, plotted in Fig. 7, indicate that in the local Universe ($0.001 < z < 0.1$), the direction of bulk flow aligns approximately with the CMB dipole direction ($ (l, b) = (276, 30) $). However, beyond the local universe, the direction of bulk flow diverges from the direction of the dark energy dipole.

In the second scenario, at low redshifts, the direction of bulk flow aligns with the CMB dipole, and on a large scale, it broadly agrees with the direction of the dark energy dipole (Fig. 8). Redshift tomography was employed for more precise results, considering two regions: $z < 0.1$ and $0.1 < z < 1.4$. Figs. (9, 10) illustrate the direction and amplitude of bulk velocity in specific redshift shells. The interplay between neutrinos becoming non-relativistic, transferring energy to the scalar field, and the subsequent evolution is evident, as described in Figure 6.

By conducting a meticulous analysis of the neutrino density-to-redshift ratio in Figures 2 to 5, it is inferred that as neutrino density diminishes, their gravitational influence weakens. Conversely, an increase in neutrino density corresponds to an increment in the bulk velocity. Within the redshift range of $0.1 < z < 1$, there is a notable augmentation in the bulk flow velocity, corroborated by Figures 10 and 11. In the redshift range of $1 < z < 1.4$, intriguing observations unveil a simultaneous reduction in neutrino density and the amplitude of bulk velocity. Notably, the bulk flow direction is aligned with the dark energy dipole's direction within this specific redshift interval. This alignment prompts the exploration of a potential correlation between the behavior of neutrinos and the dynamic nature of dark energy.  The observed decline in neutrino density appears to correspond with a parallel decrease in the amplitude of bulk velocity. This dual phenomenon, occurring within the specified redshift range, raises compelling questions about the interplay between neutrinos and dark energy dynamics. The alignment of the bulk flow direction with the dark energy dipole direction further emphasizes the nuanced relationship  within this cosmic epoch.

These results suggest that non-relativistic neutrinos may contribute to the anisotropy in the Universe and could be a factor influencing the dark energy dipole and the CMB dipole.

\end{document}